\documentclass[preprint,5p,times,twocolumn]{elsarticle}

\usepackage{amssymb}
\usepackage{amsmath}
\usepackage{mathtools}
\usepackage{dsfont}
\usepackage{pifont}
\usepackage{booktabs}
\usepackage{array}
\usepackage{tabularx}
\usepackage{enumitem}
\usepackage{stfloats}
\usepackage{subcaption}
\usepackage{caption}
\usepackage{listings}
\usepackage{xurl} 
\usepackage{float}
\usepackage{placeins} 

\lstdefinelanguage{YAML}{
  keywords={true,false,null,yes,no,on,off},
  keywordstyle=\color{blue}\bfseries,
  comment=[l]{\#},
  commentstyle=\color{gray}\itshape,
  stringstyle=,
  morestring=[b]',
  morestring=[b]",
  sensitive=true,
  columns=fullflexible,
}

\journal{Future Generation Computer Systems}

\begin{document}

\begin{frontmatter}



\title{A User-centric Kubernetes-based Architecture for Green Cloud Computing}


\author[1]{Matteo Zanotto} 
\author[1]{Leonardo Vicentini}
\author[1]{Redi Vreto}
\author[2]{Francesco Lumpp}
\author[2]{Diego Braga}
\author[1]{Sandro Fiore}

\affiliation[1]{
    organization={University of Trento},
    city={Trento},
    state={Italy}
}
\affiliation[2]{
    organization={Krateo},
    city={Verona},
    state={Italy}
}

\ead{sandro.fiore@unitn.it}

\begin{abstract}

To meet the increasing demand for cloud computing services, the scale and number of data centers keeps increasing worldwide. This growth comes at the cost of increased electricity consumption, which directly correlates to CO\textsubscript{2} emissions -- the main driver of climate change. As such, researching ways to reduce cloud computing emissions is more relevant than ever. However, although cloud providers are reportedly already working near optimal power efficiency, they fail in providing precise sustainability reporting. This calls for further improvements on the cloud computing consumer's side. 
To this end, in this paper we propose a user-centric, Kubernetes-based architecture for green cloud computing. We implement a carbon intensity forecaster and we use it to schedule workloads based on the availability of green energy, exploiting both regional and temporal variations to minimize emissions. 
We evaluate our system using real-world traces of cloud workloads execution comparing the achieved carbon emission savings against a baseline round-robin scheduler. Our findings indicate that our system can achieve up to a 13\% reduction in emissions in a strict scenario with heavy limitations on the available resources.

\end{abstract}




\end{frontmatter}




\section{Introduction}
\label{introduction}

In the past decades, cloud computing has become the backbone technology of the modern digital infrastructure, as it enables scalable and flexible deployment of heterogeneous services across globally distributed data centers. However, to meet the demand for computing resources, the scale and number of data centers have been continuously increasing.
This growth results in relevant increases in electricity demand. Today, data centers are estimated to consume 1\% of the globally available energy, and it is estimated that this usage will increase up to 13\% by 2030 \cite{masanet2020recalibrating}. Furthermore, energy consumption is one of the main contributors to the emissions of CO\textsubscript{2}, which drives climate change due to its contribution to the greenhouse effect. Currently, data centers account for 0.3\% of global emissions \cite{jones2018stop}. Since this figure is projected to increase along with the growing electricity consumption, it is imperative to research sustainability solutions to reduce the carbon footprint in cloud computing environments.

Lowering carbon emissions requires using greener energy and reducing overall electricity consumption. Cloud providers have reportedly moved in both directions.
Green energy usage has been increased with purchase agreements for Renewable Energy Certificates (RECs) and by deploying additional on-site generators of renewable energy \cite{GoogleSustainabilityRecs} \cite{GoogleDataCenters2025} \cite{AmazonSustainability2025} \cite{MicrosoftAzure2025}. Instead, electricity consumption has been reduced by upgrading physical infrastructures with more power-efficient components, such as processors with lower idle power usage and data storage devices that are denser and more energy-efficient. Moreover, increased virtualization usage greatly reduced the number of required machines by packing multiple computing instances in the same server. Overall, the impact of these improvements is reflected in the steady decrease in the Power Usage Effectiveness (PUE) of data centers as reported by cloud providers \cite{masanet2020recalibrating}.

Still, with the proliferation of cloud computing for Artificial Intelligence services based on Large Language Models (LLMs), Software-as-a-Service (SaaS) applications, Internet of Things (IoT) devices, and content distribution, cloud services demand is projected to keep increasing \cite{masanet2020recalibrating}. As such, energy consumption and its corresponding carbon footprint remain relevant sources of concern for the cloud computing industry. Thus, further advancements are required to mitigate their impact. Furthermore, despite the measures that cloud providers reportedly adopted to improve their energy efficiency, such as servers upgrades and green energy generation, they fail to provide precise reporting of sustainability metrics, such as the electricity consumption and carbon footprint of specific workloads or the embodied emissions of hardware components \cite{LBNL2024} \cite{mytton2020assessing} \cite{mytton2020hiding}. This lack of transparency reduces the trustworthiness of cloud providers regarding their sustainability and calls for improvements also to be implemented on the cloud consumers' side with software-based solutions. 

To this end, aligning the scheduling of workloads with the availability of green energy has emerged as a promising approach. The amount of carbon emissions associated with electricity consumption varies greatly depending on the sources used to power the underlying grid. Regional and temporal patterns can be observed depending on the availability of green energy and the technologies employed for generation. For example, solar power can be harnessed in regions with rich solar exposure, but is available only during daylight hours; hydroelectric power instead, is abundant in regions with ample water resources but exhibits seasonal variations based on precipitation patterns. The carbon intensity of the underlying power grid will thus be different at varying locations and times of the day.

Carbon intensity variations can be exploited to optimize the scheduling of cloud workloads to minimize their carbon footprint. Many cloud workloads present temporal and spatial tolerance. Non-critical jobs can be delayed up to a final deadline, for example, in the case of integration testing executed overnight before the next day. For non-interactive jobs that do not require timely responses to users, such as the training of machine learning models, their insensitivity to latency can be leveraged to relocate them to areas with a lower carbon intensity.

To enable time-based scheduling of workloads according to green energy availability, it is crucial to forecast future carbon intensity levels. Due to their time-related nature, carbon intensity levels can be represented as time series data. Advances in Machine Learning techniques have shown that methods such as Deep Learning are promising approaches to time series forecasting \cite{lim2021time}, even outperforming classical statistical methods \cite{kontopoulou2023review} \cite{siami2018comparison}. Therefore, a dedicated carbon intensity forecasting model using Machine Learning can effectively support the dynamic scheduling of cloud workloads.

In this paper, we propose a Kubernetes-based architecture for the management of cloud workloads to reduce their carbon footprint. More specifically, we adopt a user-centric approach by considering information and APIs available on the cloud computing consumers' side. In doing so, both approaches of space- and time-shifting scheduling are used to identify the best region and period in which to allocate workloads. This is enabled by the implementation of a forecaster component that can generate future carbon intensity levels at geographically distributed regions.
To summarize, the main contributions of this paper are:
\begin{itemize}
\item A comprehensive scheduling approach that considers the combined effects of both spatial and temporal workload shifting to optimize carbon footprint.
\item A practical implementation that addresses limitations in prior research by focusing on consumer-level cloud resource allocation rather than infrastructure-level solutions.
\item A carbon-aware scheduler for cloud workloads implemented as an open-source Kubernetes-based architecture with carbon intensity forecasting.
\end{itemize}

The rest of this paper is structured as follows. Section \ref{relatedwork} analyzes other relevant architectures from state-of-the-art studies. Background information about the concepts and technologies used in this study is then presented in Section \ref{background}. Our formulation of sustainable cloud computing scheduling is presented in Section \ref{problemstatement}, while Section \ref{system_design_and_implementation} describes the implementation of our architecture. The experimental evaluation of our system is presented in Section \ref{evaluation}. Finally, Section \ref{conclusion} wraps up the paper and provides insights for future developments.


\section{Related work}
\label{relatedwork}

Reducing the carbon footprint of computation has become a problem of great research interest due to its increasingly severe impact. In the context of cloud computing, aligning the schedule of workloads with the availability of green energy has proved to be a promising approach for minimizing the amount of greenhouse gases associated with the execution of software. In previous research, various architectures have been proposed for the carbon-aware scheduling of cloud workloads, using both geographical (space-shift scheduling) and temporal (time-shift scheduling) variations.

\subsection{Space-shift scheduling}
\label{spaceshift}

The minimization of the carbon footprint in cloud workload scheduling through offloading across geographically distributed clusters is referred to as "space shifting" scheduling, and it has been extensively studied.

Various optimization models have been proposed, often relying on heuristic techniques for practical solutions. Garg et al. \cite{garg2011environment} introduced a set of heuristic policies to schedule HPC cloud workloads between multiple data centers, achieving a trade-off between minimizing carbon emissions and maximizing profit. Doyle et. al \cite{doyle2013stratus} used Voronoi cells to assign requests to various data centers to minimize latency, carbon footprint, and electricity costs. Kaur et al. \cite{kaur2019multi} formalized a multi-cluster scheduler composed of a workload classifier and a multi-objective problem to achieve a trade-off between carbon emissions, Service Level Agreements (SLA) agreements, and energy costs. Xu and Buyya \cite{xu2020managing} proposed the specification of a multi-cloud workload scheduling algorithm designed to maximize renewable energy utilization through the brownout mechanism, which selectively disables optional application features to enhance sustainability. 

Practical implementations of these models have also been explored. James et al. \cite{james2019low} presented a Kubernetes scheduler extension in a multi-cloud setting that accounts for the local air temperature at data centers. Similarly, Kaur et al. \cite{kaur2019keids} introduced KEIDS, a Kubernetes system for scheduling IoT jobs in a multi-cloud setting that achieves optimal server consolidation while avoiding CPU interference.

The benefits of offloading cloud workloads between geographically distributed data centers are greatly reduced when there are strict latency limits \cite{sukprasert2024limitations}. This is the case for interactive services that require timely responses to the users. Despite these challenges, spatially distributed workload scheduling has also been investigated in the context of web services and the Function-as-a-Service (FaaS) paradigm. Toosi et al. \cite{toosi2017renewable} proposed a weighted load-balancer that redirects web requests to multiple data centers to maximize renewable energy usage. Souza et al. \cite{souza2023casper} implemented Casper, a Kubernetes scheduler extension for load-balancing web requests between various data centers. Murillo et al. \cite{murillo2024cdn} described a load-balancing algorithm for managing Content Delivery Networks (CDNs) with a tradeoff between emissions, costs, and latency with long-term migration of server capacity to greener regions. Chadha et al. \cite{chadha2023greencourier} proposed a Kubernetes scheduler extension called GreenCourier for distributing serverless functions across multiple clusters. Gsteiger et al. \cite{gsteiger2024caribou} implemented Caribou, a Python middleware library for the automatic fine-grained deployment of serverless functions across different regions.

\subsection{Time-shift scheduling}
\label{timeshift}

Minimizing the carbon footprint of cloud workloads by scheduling them to periods of ample green energy availability is referred to as "time shifting" scheduling. Various studies have evaluated its benefits in synchronizing jobs with low carbon times. Goiri et al. \cite{goiri2011greenslot} implemented GreenSlot, a greedy workload scheduler that maximizes renewable energy usage based on solar power forecasts. Radovanovic et al. \cite{radovanovic2022carbon} presented the Virtual Capacity Curves (VCCs) used within Google’s data centers to artificially limit the available computing resources in periods of high emissions, delaying non-critical jobs. Weisner et al. \cite{wiesner2021let} proposed the Wait Awhile algorithm to quantify the carbon emissions savings achieved by delaying jobs to low-emissions periods. Bostandoost et al. \cite{bostandoost2024data} proposed a meta-algorithm to apply the best time-shifting heuristic depending on the characteristics of the workload. 

An example of a practical implementation of a time-shifting scheduler is proposed by Hanafy et al \cite{hanafy2024going} as they describe an AWS Slurm extension to delay jobs to minimize emissions with a trade-off with cloud resources allocations costs.
Time-shifting scheduling has also been integrated into Kubernetes-based cloud environments. Hanafy et al. \cite{hanafy2023carbonscaler} proposed CarbonScaler, a Kubernetes operator that controls the execution times of workloads by dynamically scaling their computational resources to match green energy periods. Similarly, Piontek et al. \cite{piontek2024carbon} extended the default Kubernetes scheduler to delay jobs to low-carbon periods based on carbon intensity forecasts.

\subsection{Space- and time-shift scheduling}
\label{timespaceshift}

Few studies have explored the combined approach of time- and space-shifting for scheduling workloads across different times and geographic regions. Le et al. \cite{le2015scheduling} evaluated a set of greedy heuristics for the Dijkstra's algorithm for scheduling workloads in a multi-cluster setting with a look-ahead time window. Zhao et al. \cite{zhao2022energy} proposed a virtual machine (VM) placement algorithm that maximizes the usage of renewable energy at distributed clusters by predicting future wind intensity. Bahreini et al. \cite{bahreini2023carbon} formalized a scheduling algorithm to dispatch workloads in regions and times with minimal carbon intensity. This algorithm was later used to implement Caspian \cite{bahreini2024caspian}, a Kubernetes controller that schedules jobs in regions and times with the lowest carbon intensity. Finally, although only prototyped, Cla\ss en et al. \cite{classen2023carbon} developed a carbon-aware scheduler that selects the optimal time and region to minimize emissions of CI/CD jobs.

\subsection{Limitations and research gap}
\label{limitations}

Previous research has explored both temporal and geographical workload schedule shifting as strategies for minimizing the carbon footprint of cloud computing. Various systems have been proposed to align cloud usage with the availability of green energy at varying times of the day and in different regions of the world. However, they all have certain limitations in some respects, which we aimed at addressing in this study. 

First of all, many of the previous studies address the problem of sustainable cloud scheduling primarily from a modeling perspective within simulated environments \cite{garg2011environment} \cite{doyle2013stratus} \cite{le2015scheduling} \cite{kaur2019multi} \cite{xu2020managing} \cite{wiesner2021let} \cite{zhao2022energy} \cite{bahreini2023carbon} \cite{bostandoost2024data} \cite{murillo2024cdn}. While useful for theoretical exploration, this approach often omits critical implementation details of the underlying architecture and relies on assumptions that may not hold in real-world systems. For example, many systems compute the optimal jobs schedule using information about the power consumption of workloads and the carbon intensity of the underlying power grid. In practice, this data is available only by usage of third-party commercial services such as Electricity Maps \cite{em_website} or WattTime \cite{watttime_website}, or by the implementation of specific forecasters.

Indeed, another limitation of previous research is that very few architectures \cite{goiri2011greenslot} \cite{zhao2022energy} include an integrated forecaster to predict future carbon intensity or the availability of green energy. Most existing studies instead rely on historical data to evaluate their architecture, offering only a conceptual abstraction for how real-time data would be retrieved during the online operation of the carbon-aware scheduler. However, this component is essential when the scheduling system implements time-shifted schedules as it needs to find the best future window to delay jobs to minimize emissions. Therefore, when designing a practical and adoptable architecture, the inclusion of a forecasting component is crucial.

Another major limitation of existing research is its limited applicability from the perspective of cloud consumers. This is particularly significant given that cloud providers are already operating near optimal efficiency, making consumer-side optimization crucial for further emissions reductions. This limitation is imposed by optimizing workloads schedule using parameters tied to the internal states of data centers, which are inaccessible to external users. For example, some systems depend on controlling hardware-level mechanisms \cite{garg2011environment} \cite{toosi2017renewable} while others rely on internal metrics like aggregate resource utilization across tenants \cite{toosi2017renewable}\cite{radovanovic2022carbon} or on-site renewable energy availability \cite{kaur2019multi} \cite{zhao2022energy} \cite{goiri2011greenslot}. Such parameters are generally unavailable to consumers, who interact with cloud infrastructure only through high-level APIs that abstract away physical and infrastructural details.

Furthermore, while only a few studies \cite{james2019low} \cite{kaur2019keids} \cite{piontek2024carbon} \cite{souza2023casper} \cite{hanafy2023carbonscaler} \cite{chadha2023greencourier} provide a practical implementation based on Kubernetes, most of the other proposed approaches do not. This is essential as Kubernetes has become the standard for managing cloud-native workloads, and the lack of a compatible implementation significantly limits the applicability and adoption of many solutions by cloud computing practitioners.

Similarly, although some studies propose systems that can be easily extended for client-specific policies and constraints \cite{garg2011environment} \cite{classen2023carbon} \cite{james2019low} \cite{souza2023casper} \cite{hanafy2024going}, these capabilities are not widely supported across the solutions proposed in the literature. Such flexibility is crucial given the diversity of cloud services, SLA, and Quality of Service (QoS) requirements. For example, enforcing data residency constraints, to limit processing to specific regions to comply with the General Data Protection Regulation (GDPR), demands dynamic updates to the scheduling formulation when offloading certain kinds of critical workloads geographically. As such, the lack of adaptability in the proposed systems further reduces their adoption by cloud computing users.

Table \ref{tab:related_work_comparison} summarizes the state of the art and the main differences with this work. The first two columns show whether the work makes use of space-shift and time-shift techniques, respectively. The third column indicates if the work has practical implementations or if its a modeled solution. The fourth column focuses on the target audience of the proposed approach, either the user of the infrastructure or the maintainer. Whether the solution is Kubernetes-based is indicated in the fifth column. The sixth column shows the support for client policies. Finally, the seventh column indicates if the proposed work implements a carbon forecaster.

To address the limitations identified in prior research, we propose a flexible, user-centric carbon-aware scheduler for cloud workloads. More specifically, the proposed design targets cloud computing consumers who seek to minimize the carbon footprint of the cloud resources allocated through interactions with the high-level APIs of cloud providers. To this end, we implemented a carbon-aware scheduler as a Kubernetes-based architecture comprising a carbon intensity forecaster. In this way, we aim at maximizing the adoption of the proposed system by adherence to industry standards while minimizing the carbon footprint of cloud workloads by considering the combined effects of both space- and time-shifted schedules.

\begin{table*}[ht] 
\centering
\renewcommand{\arraystretch}{1.2}
\resizebox{\textwidth}{!}{%
\begin{tabular}{>{\raggedright\arraybackslash}p{3.5cm} 
                >{\centering\arraybackslash}p{1.5cm}
                >{\centering\arraybackslash}p{1.5cm}
                >{\centering\arraybackslash}p{2cm}
                >{\centering\arraybackslash}p{2cm}
                >{\centering\arraybackslash}p{2cm}
                >{\centering\arraybackslash}p{2cm}
                >{\centering\arraybackslash}p{3cm}}
\toprule
\textbf{Study} & \textbf{Space-shift} & \textbf{Time-shift} & \textbf{Practical impl.} & \textbf{User-centric} & \textbf{Kubernetes-based} & \textbf{Client policy support} & \textbf{Carbon forecaster}  \\
\midrule

Garg et al.\cite{garg2011environment} & \ding{51} & & & & & & \ding{51} \\

Doyle et. al\cite{doyle2013stratus} & \ding{51} & & & & & & \\

Kaur et al.\cite{kaur2019multi} & \ding{51} & & & & & & \\

Xu and Buyya\cite{xu2020managing} & \ding{51} & & & & & & \\

James et al.\cite{james2019low} & \ding{51} & & \ding{51} & \ding{51} & \ding{51} & & \\

Kaur et al.\cite{kaur2019keids} & \ding{51} & & \ding{51} & & \ding{51} & & \\

Toosi et al.\cite{toosi2017renewable} & \ding{51} & & \ding{51} & & & & \\

Souza et al.\cite{souza2023casper} & \ding{51} & & \ding{51} & & \ding{51} & & \\

Murillo et al.\cite{murillo2024cdn} & \ding{51} & & & & & & \\

Chadha et al.\cite{chadha2023greencourier} & & \ding{51} & \ding{51} & \ding{51} & \ding{51} & \ding{51} & \\

Gsteiger et al.\cite{gsteiger2024caribou} & \ding{51} & & \ding{51} & \ding{51} & & \ding{51} & \\

Goiri et al.\cite{goiri2011greenslot} & & \ding{51} & \ding{51} & & & & \ding{51} \\

Radovanovic et al.\cite{radovanovic2022carbon} & & \ding{51} & & & & & \\

Weisner et al.\cite{wiesner2021let} & & \ding{51} & & & & & \\

Bostandoost et al.\cite{bostandoost2024data} & & \ding{51} & \ding{51} & \ding{51} & & & \\

Hanafy et al.\cite{hanafy2024going} & & \ding{51} & \ding{51} & \ding{51} & & \ding{51} & \\

Hanafy et al.\cite{hanafy2023carbonscaler} & & \ding{51} & \ding{51} & \ding{51} & \ding{51} & & \\

Piontek et al.\cite{piontek2024carbon} & & \ding{51} & \ding{51} & \ding{51} & \ding{51} & & \\

Le et al.\cite{le2015scheduling} & \ding{51} & \ding{51} & & & & & \\

Zhao et al.\cite{zhao2022energy} & \ding{51} & \ding{51} & & & & & \ding{51} \\

Bahreini et al.\cite{bahreini2023carbon} & \ding{51} & \ding{51} & & & & & \\

Bahreini et al.\cite{bahreini2024caspian} & \ding{51} & \ding{51} & \ding{51} & \ding{51} & \ding{51} & & \\

Cla\ss en et al.\cite{classen2023carbon} & \ding{51} & \ding{51} & \ding{51} & \ding{51} &  & \ding{51} & \\

\textbf{Our Approach} & \ding{51} & \ding{51}   & \ding{51} & \ding{51}   & \ding{51}   & \ding{51} & \ding{51}   \\

\bottomrule
\end{tabular}%
}
\caption{Comparison of related architectures for sustainable cloud computing.}
\label{tab:related_work_comparison}
\end{table*} 

\subsection{Time series forecasting}

In addition to exploring architectures for sustainable cloud computing, we evaluated several state-of-the-art models for time series forecasting to inform the design of our carbon intensity forecaster component. 

Time-LLM \cite{jin2023time} is an LLM-based framework that transforms raw time series data into textual representations for prediction. However, as reported by Tan et al. \cite{tan2024language}, the incorporation of LLMs has not consistently yielded performance improvements over traditional approaches.
Lag-Llama \cite{rasul2023lag}, based on the LLaMA architecture \cite{touvron2023llama}, provides univariate probabilistic forecasting, using lagged values as covariates.
TimesFM \cite{das2024decoder} is a decoder-only transformer model designed specifically for time series forecasting. It is both flexible and computationally efficient, but lacks support for incorporating exogenous variables.
TinyTimeMixers \cite{ekambaram2024tiny} are  compact pre-trained models designed for multivariate time-series forecasting. They outperform larger models like TimesFM with significantly lower computational demands. Notably, they support the integration of exogeneous data, enabling more comprehensive modeling of external factors.
Moment \cite{goswami2024moment} is a family of open-source foundation models designed for general-purpose time-series analysis, such as forecasting, classification or anomaly detection.
Chronos \cite{ansari2024chronos} adapts transformer-based language models for probabilistic time series forecasting, tokenizing time series data into discrete sequences.

Ultimately, we based the proposed  implementation of the carbon intensity forecaster on the 1024 context-length version of the TinyTimeMixers (TTMs) model due to its low computational requirements and superior zero-shot benchmarking performance. Furthermore, it supports the integration of exogenous data, allowing for more comprehensive modeling of external factors influencing the time series.


\section{Background}
\label{background}

This section provides some background information on the concept of carbon intensity and on the technologies that were used during this study.

\subsection{Carbon intensity}
\label{ci}

Carbon intensity (measured in $gCO_2eq/kWh$) quantifies the greenhouse gas emissions associated with electricity production and reflects the environmental impact of power consumption. It varies depending on the energy sources, with fossil fuels producing high emissions, while renewable sources such as wind or solar yield near-zero emissions. Due to the variable availability and limited storage capacity of renewables, they are used immediately when available, with additional demand met with fossil fuel-based generators. As a result, carbon intensity fluctuates over time and across regions, typically increasing during peak power demand. 

\subsection{Electricity Maps}
Electricity Maps \cite{em_website} is a service that provides historical and real-time global electricity grid data at hourly resolution. For each region, which typically corresponds to a state or an electricity provider coverage area, a CSV file reports timestamped average and marginal carbon intensity, renewable energy share, and power production and consumption measures. For the scope of this work, we accessed data from 2010 to 2022.

\subsection{Kubernetes}
\label{kubernetes}

Kubernetes \cite{kubernetes_website} has become the de facto standard for container orchestration, offering a highly extensible control plane for managing arbitrary resources, including those external to the cluster.

Its extensibility is enabled by Custom Resource Definitions (CRDs) and the operator pattern. 
CRDs allow users to define new resource types that are validated by the API server and can be dynamically registered within the cluster. The schema specified by a CRD is instantiated into a Kubernetes cluster as a Custom Resource (CR).
Operators manage the lifecycle of Kubernetes resources by implementing reconcile loops that ensure the current state of the resources matches the desired state. Custom operators can be developed to manage CRs with specific logic. Notably, operators can also interact with external APIs, enabling Kubernetes to manage any kind of out-of-cluster resource.

The Kubernetes admission control is another extensibility point as it governs how API requests are processed before being persisted in the cluster.
In addition to built-in admission plugins, Kubernetes supports custom admission control mechanisms as external webhooks, which can be dynamically registered at runtime without modifying the core API server. They can be used to enforce policies or to transform Kubernetes objects based on custom logic. Validating webhooks approve or reject objects based on policies, while mutating webhooks modify objects before they are persisted.
Webhooks are essentially servers that accept and respond to requests. As such, they can be implemented either as in-cluster services or as external endpoints that need to be accessible to the Kubernetes API server.

\subsection{Helm}

Helm\footnote{https://helm.sh/docs/} is a widely adopted package manager for Kubernetes that simplifies the definition, deployment, and management of applications. It introduces the concept of charts, which are parametrized templates to collect and configure a set of related Kubernetes resources. Upon chart installation, Helm renders these templates by injecting the variable values and deploying the resources in the Kubernetes cluster. This mechanism enhances the customization and reproducibility of Kubernetes resources while reducing the operational overhead in their management.

\subsection{Open Policy Agent}
\label{opa}

Open Policy Agent (OPA)\footnote{https://www.openpolicyagent.org/docs} is an open-source, general-purpose policy engine that enables unified policy decision-making across several types of environments. 
Policies are declaratively defined using the Rego high-level language, enabling the ``Policy as Code" paradigm. As such, policies and associated contextual data can be bundled and distributed as versioned packages, ensuring centralized and consistent policy management and maintainability across distributed environments.
OPA  can be integrated as a sidecar container, host-level daemon, or as a library and is widely used for use cases such as Kubernetes admission control, CI/CD pipelines, and container image validation.


\subsection{Krateo}
\label{krateo}

Krateo\footnote{https://docs.krateo.io/} is an open-source Kubernetes-based platform that provides a unified control plane for managing arbitrary resources, both internal and external to the cluster. 
It operates as a Kubernetes deployment, and it handles the declarative management of resources represented as Helm charts. 

Its main component, the core-provider, consists of a Kubernetes operator that automates the lifecycle management of Helm charts within Kubernetes, which does not provide a way of managing Helm charts out of the box. 
This is done by introducing the \textit{CompositionDefinition}, which is a CRD that references a Helm chart template and its JSON Schema.





Upon applying a CompositionDefinition CR to the Kubernetes cluster, the core-provider generates an additional CRD representing the possible values that can be inserted in the Helm chart it specifies according to the associated JSON schema. This mechanism ensures validation of the Helm charts before installation, as this CRD will be used to install and manage Helm charts within the cluster.

For each CompositionDefinition CR received by the core-provider, it deploys a dedicated instance of the \textit{composition-dynamic-controller} (CDC). It is a custom controller responsible for the reconcile loop of the CRDs representing the possible Helm chart values.

In this way, once a CR instance of these CRDs is applied to the cluster, the dedicated CDC first checks for the existence of the corresponding Helm release. If absent, it performs a Helm install with the values specified in the CR, while if it already exists, it performs a Helm upgrade with the new values. When the resource is deleted, the CDC performs a Helm uninstall on the release.

This mechanism ensures full lifecycle support of Helm charts within Kubernetes, enabling Krateo to manage heterogeneous resources in a declarative manner. 
This feature will be leveraged for the instantiation of cloud provider-specific virtual machine resources and for the scheduling wait time logic, as discussed in Section \ref{sec:wait-logic}.


\section{Problem statement}
\label{problemstatement}

We addressed the problem of carbon-aware cloud computing considering a user-centric approach in a multi-cluster setting. More specifically, we tackled the use case in which cloud computing consumers need to minimize the carbon footprint of the workloads they allocate through interaction with the high-level APIs of cloud providers, who provide computing resources hosted at geographically distributed data centers. To this end, our architecture acts as an intermediary as it receives the requests and determines the optimal schedule. This is done by considering both geographical offloading and temporal delaying of jobs, while guaranteeing SLA such as the amount of requested resources (i.e., CPU and RAM) and deadlines. By adopting a user-centric approach, the optimization is performed without considering information inaccessible to cloud consumers in real-world scenarios, meaning it does not require data about the inner states of computing clusters. Instead, our system relies on carbon intensity forecasts produced by its forecaster component (described in Section \ref{forecaster}). Moreover, the scheduled resources are allocated through the same high-level APIs that cloud providers expose to users. In this way, our architecture sits on the consumer's side, providing a ``sustainability layer'' that can be easily adopted over the user's already existing cloud-native infrastructure.

We modeled the workloads whose allocation is requested by users as VMs. We used this representation as VMs are a common and widely used cloud resource that can be easily provisioned on multiple cloud providers. Moreover, they are a key infrastructural component as they host services and applications. We consider a VM to be represented as a tuple:

\[
\text{VM} \coloneqq (\text{MinCPU}, \text{MinRAM}, D, DL, ML)
\]

where:
\begin{itemize}[itemsep=0.2pt, topsep=1pt]
    \item \( \text{MinCPU} \) is the minimum number of virtual CPUs required.
    \item \( \text{MinRAM} \) is the minimum RAM required (in GB).
    \item \( D \) is the duration for which the VM must run to complete its arbitrary processing task (in hours).
    \item \( DL \) is the deadline timestamp by which the VM must complete execution.
    \item \( ML \) is the maximum allowed latency (in milliseconds).
\end{itemize}

For the scope of this work, we assume workloads to be non-preemptible, i.e., once a VM is allocated, it cannot be stopped for its whole duration. Furthermore, we assume that the resources allocated to VMs are not scaled during their lifetime, meaning no CPU cores or RAM amounts are dynamically added or removed while the VM is running.


To determine the optimal schedule for the incoming VM request, namely its allocation time and region, the scheduler solves an optimization problem. More specifically, a binary integer programming formulation is used over a discretely time-slotted horizon of hourly resolution. Given the VM $k$ to allocate as the tuple $\text{VM}_k = (\text{MinCPU}_k, \text{MinRAM}_k, D_k, DL_k, ML_k)$, let $x_{tjk}$ be a binary variable such that:
\[
x_{tjk} = \left\{
\begin{array}{ll}
1, & \text{if job } k \text{ is allocated at datacenter } j \text{ at time } t, \\
0, & \text{otherwise}.
\end{array}
\right.
\]
with $c_{tjk}$ being the cost in total carbon intensity of the allocation, given by:
\[
c_{tjk} = \sum_t^{t+D_k} CI_j[t]
\]
where $D_k$ is the duration in hours of VM $k$ and $CI_j[t]$ is the carbon intensity at time $t$ in region $j$. It is to be noted that this cost does not directly correspond to the carbon emissions of the workload as it depends on the power consumption profile of the workload. However, it is reasonable to assume without any loss of accuracy a constant power consumption corresponding to a fixed average resource usage at each instant. This approximation will result in the same overall consumption, only smoothing the various spikes that would happen in a real-world scenario. Additionally, since the average will amount to a constant in the summation, it can be omitted.

Using this definition, the optimal schedule of $K$ jobs to be allocated over $J$ regions over a look-ahead window of length $T = (DL_k - D_k)$ is given by:
$$
\text{min}_{x\in X} \sum_t^T\sum_j^J\sum_k^K x_{tjk}\cdot c_{tjk}
$$
subject to constraints:
\begin{enumerate}
    \item $\forall k\in K, \sum_t^T\sum_j^J x_{tjk}=1$, meaning jobs are allocated exactly once at a certain time and region.
    \item Jobs are completed before their deadline, already enforced by considering a look-ahead window of length $T = DL_k - D_k$.
    \item $\forall j\in J, \forall t\in T \sum_{k}^K x_{tjk}+\text{max}_{\tau\in[t, t+D_k]}(alloc[j][\tau])\leq M_j$ meaning no more than $M_j$ workloads are allocated in region $j$, where $alloc[j][t]$ is a matrix that counts the number of jobs running at region $j$ at instant $t$, updated on the result of the optimization. More specifically, once the optimal schedule $(j^*, t^*)$ is determined for a VM, the matrix is updated as $alloc[j][t] = alloc[j][t] + 1$ for $t\in [t^*, t^*+D_k]$.
\end{enumerate}

Constraint (3) is put in place to prevent a ``black hole'' corner case in which the optimization schedules all jobs to the region with the lowest overall carbon intensity. In practice, this would be unfeasible due to resource limitations, as a single region would be unable to handle such a high volume of requests. This is because, although abstracted by definition, the cloud computing paradigm ultimately relies on physical machines hosted in data centers, which have finite resource capacities. Moreover, the increased usage of computing resources would result in increased electricity demand, typically met by the activation of peaking power plants. These plants are often fossil-fuel based and would consequently raise the carbon intensity of the grid. To prevent this issue, we enforce a maximum of $M_j$ jobs to be allocated at region $j$. The value of $M_j$ must be determined experimentally, based on the incoming request rate and available resources. We leave the estimation of this parameter for future work.

When scheduling VMs, maximum latency is not considered during the optimization as eligible regions are already determined in advance by our policy-enforcing component, as described in section \ref{opa_integration}. It leverages latency tables containing entries for each region covered by cloud providers to report the maximum expected latency that a service allocated at a certain region will have when requests originate from another region. By locating the region from which the user submits its requests, it is possible to filter out those regions that already exceed the maximum latency beforehand. Still, periodical updates to the latency tables are required. This is done using publicly released data from cloud providers and ping tests using the cloudping tool\footnote{https://www.cloudping.info/}.

With the resulting formulation, the carbon-aware scheduling of VMs corresponds to a form of the optimal job scheduling problem. To resolve its binary integer programming optimization, we use the simplex algorithm leveraging the Python PuLP library with the CBC solver \cite{mitchell_pulp_2011} \cite{coinor_cbc_2024}.

\section{System design and implementation}
\label{system_design_and_implementation}

To address the carbon-aware scheduling problem from a user-centric perspective, we designed a Kubernetes-based architecture to manage the VM allocation requests. The proposed system builds upon the Krateo platform, integrating it with Open Policy Agent (OPA) and cloud providers operators to embed the scheduling in the Kubernetes resource mutation mechanism. Further details are provided in the following sections.

\begin{figure*}[hbt]
  \centering
  \includegraphics[width=\textwidth]{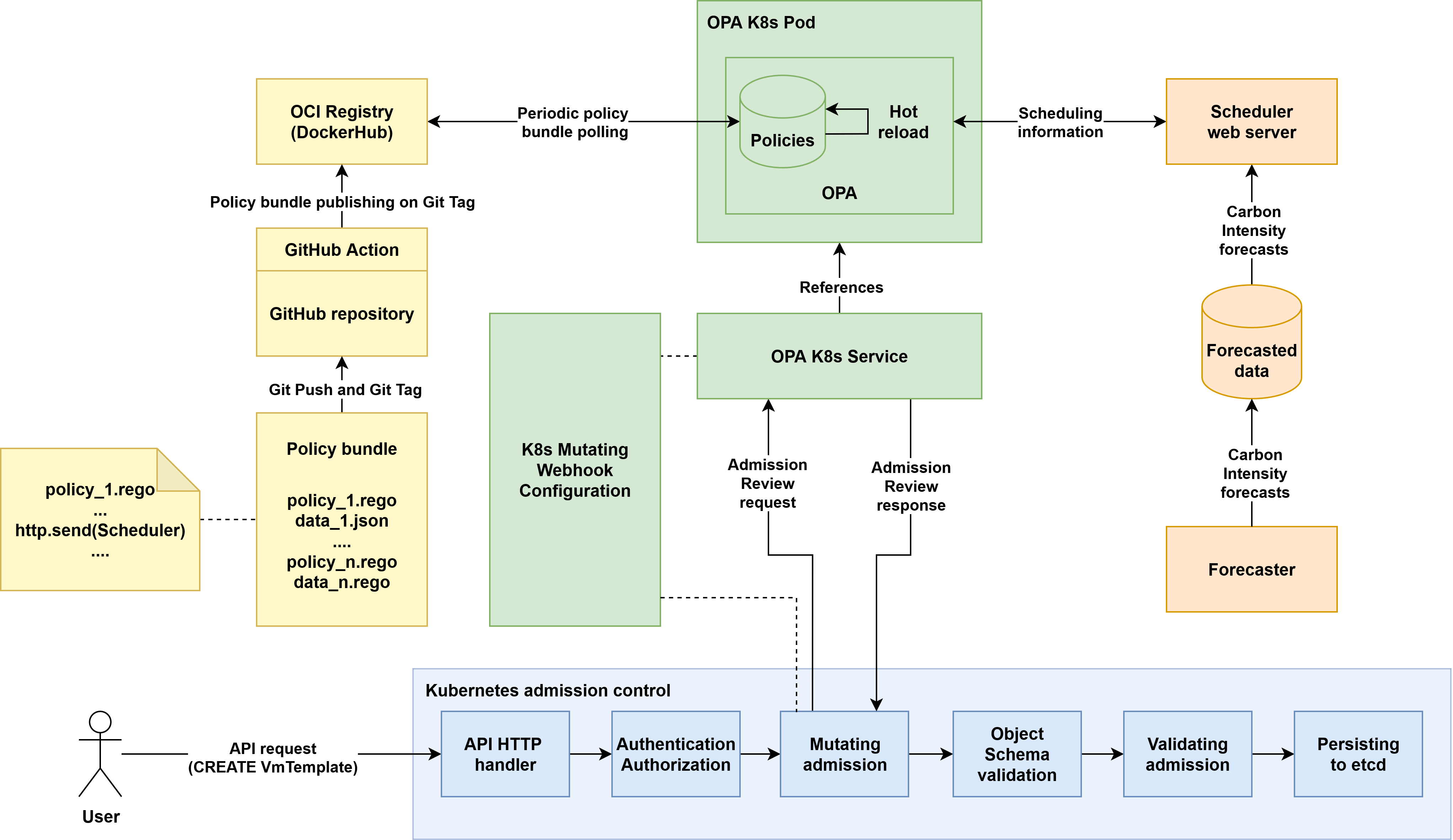}
  \caption{High-level architecture of the system. Blue components represent the Kubernetes admission control mechanism. Green components refer to the Open Policy Agent integration within the webhook configuration. Yellow components represent the management of policy bundles as code. Orange components represent the scheduler and forecaster mechanisms.}
  \label{fig:high_level_architecture}
\end{figure*}

Figure \ref{fig:high_level_architecture} represents the high-level architecture of the proposed system.
Blue components are a representation of the Kubernetes admission control mechanism (mainly within the API server) that intercepts requests for VM scheduling.
Green components are related to the Open Policy Agent integration within a mutating webhook configuration.
Yellow components represent the management of Open Policy Agent policy bundles as code leveraging GitHub.
Orange components represent the scheduler and the forecaster mechanisms of the proposed system.

\subsection{Generic VM to cloud provider specific VM mapping}
\label{sec:generic-to-specific}

To implement the generic VM as the tuple in section \ref{problemstatement}, a Custom Resource called \textit{VmTemplate} is defined as a \textit{CompositionDefinition} resource. In this way, the Krateo CDC is able to reference the Helm chart of the VmTemplate. This enables the system to leverage the Helm templating mechanism to generate cloud provider-specific resources from the generic VmTemplate resource. The generic resource contains all the cloud provider-specific sets of templates necessary for VM provisioning on their platforms. Then, only one set of templates is used at a time by the CDC according to the cloud provider specified in the VmTemplate resource. This ``brokering mechanism'', which effectively consists of routing through Helm templates, is implemented using ``guards'' in the Helm templates as reported in listing \ref{lst:guards}. 
In this way, a cloud resource specific to a cloud provider is created from a VmTemplate generic resource only if it specifies the same cloud provider. For this mechanism to properly work, we validate the generic VM CR to contain a ``provider'' field.
This flexibility allows the system to be cloud-agnostic and to support multiple cloud providers as soon as the cloud provider-specific templates are available in the Helm chart.

\begin{lstlisting}[language=YAML, caption={Helm Template guards example}, label={lst:guards}, float=b]
{{ if hasKey .Values "provider" }}
{{ $provider := .Values.provider }}
{{ if eq $provider "azure" }}
...
apiVersion: compute.azure.com/v1api20220301
kind: VirtualMachine
...
{{ else if eq $provider "aws" }}
...
\end{lstlisting}


\subsection{Scheduling time waiting logic}
\label{sec:wait-logic}

The Helm template engine is also leveraged to handle the scheduling time waiting logic.
This is crucial for the proposed system, since cloud provider operators do not support scheduling time metadata in their CRDs but instead provision the resource immediately when a CR is created.
In this case, another set of ``guards'' are leveraged to effectively block manifest creation until the scheduling time is reached. Only then will Helm create and apply the manifest, and the resource will be finally provisioned by the appropriate cloud provider operator.
To this end, the CDC will periodically perform a \textit{helm upgrade} operation to trigger the brokering mechanism through the Helm chart to check whether the scheduling time is reached.

To consider the time needed for the cloud provider operator to provision the resource, additional logic could be added in the template to anticipate the scheduling time (e.g, using a \textit{provisioningTimeOffset}).

\subsection{Multi-cloud integration through Kubernetes operators}
\label{sec:cloud_providers_operators}

Integrating operators from different cloud providers enables the development of an effective multi-cloud system, allowing seamless orchestration and provisioning of cloud resources across various cloud platforms. 
The proposed system leverages Kubernetes operators from Microsoft Azure, Google Cloud Platform, and Amazon Web Services. Each operator, when installed on a Kubernetes cluster, provides a set of CRDs that represent specific cloud resources of the cloud provider, such as VMs instance types. These CRDs can be used to define cloud resources in a declarative manner as Kubernetes CRs, allowing users to specify the desired state of the cloud resources they wish to manage.
Each operator also installs a controller that monitors changes in the CRs on the cluster and takes all the necessary actions for reconciling the actual state with the desired state. Under the hood, the controller interacts with the cloud provider's API to provision, update, and delete cloud resources.
In this way, by leveraging the continuous reconciliation principle of Kubernetes resources, real-time representations of the provisioned cloud resources, as managed by the operators, are accessible within the Kubernetes cluster.

It must be noted that different cloud providers adopt different design choices for their Kubernetes operators and more in general for their overall cloud infrastructure management. 
Therefore, for the creation of logically similar resources, like virtual machines, the structure and the field of the resources can be different. 
These resources typically include:
\begin{itemize}
  \item Compute resources (e.g., VM instances, VM templates)
  \item Networking components (e.g., virtual networks, subnets, security groups)
  \item Storage allocations (e.g., persistent volumes, cloud disks)
  \item Access management (e.g., resource groups, roles, authentication credentials)
\end{itemize}
For the purpose of this work a baseline infrastructure was defined for each major cloud provider to have a common ground for the system to work. 
This baseline infrastructure is composed by the minimum set of resources needed for VM provisioning.

\begin{figure*}[htb]
  \centering
  \includegraphics[width=\textwidth]{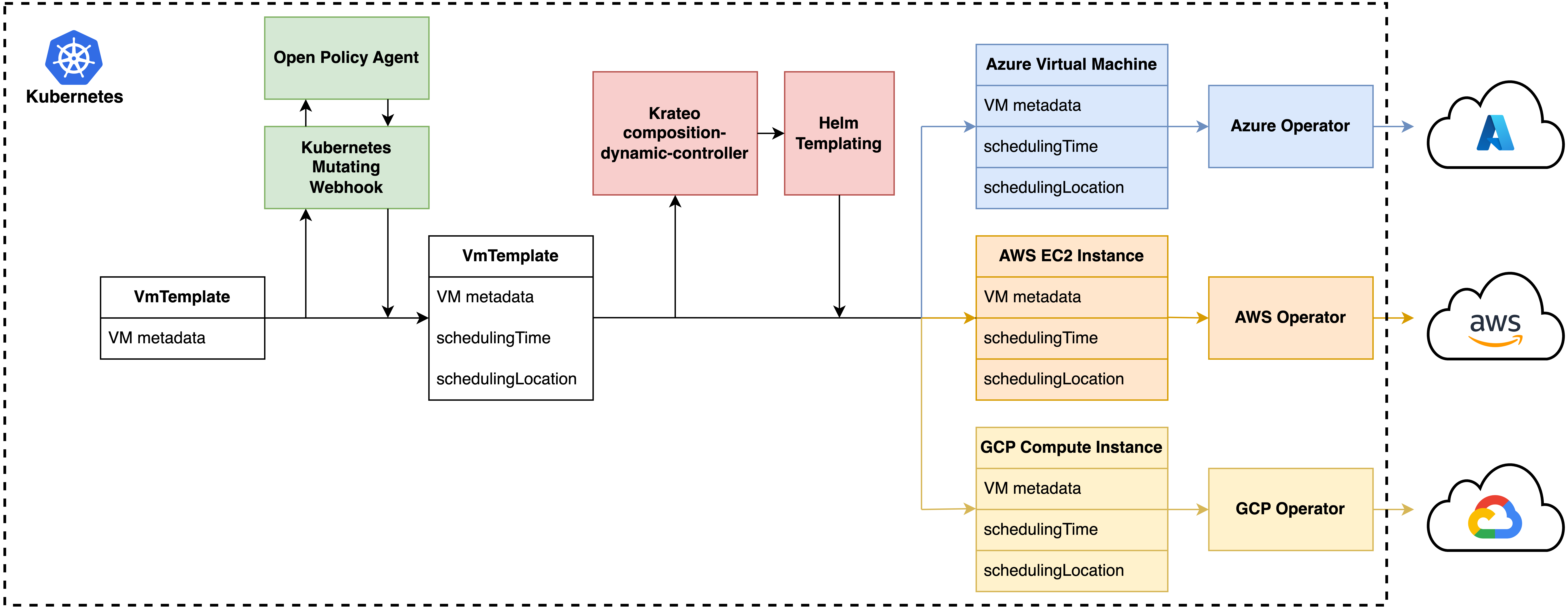}
  \caption{Multi-cloud resource management: from a generic VmTemplate representation to cloud provider-specific resources managed by dedicated operators.}
  \label{fig:generic_to_specific}
\end{figure*}

\subsection{Kubernetes mutating webhook configuration}

Within the proposed system, the scheduling process is embedded in the Kubernetes resource mutation mechanism. To this end, a \textit{MutatingWebhookConfiguration} was configured to intercept ``CREATE" and ``UPDATE" Kubernetes API requests for VmTemplate CRs to apply the scheduling logic. 
More specifically, the Kubernetes Mutating Webhook server targeted by the MutatingWebhookConfiguration is an Open Policy Agent instance, which enforces the user-defined policies and invokes the scheduler to determine the optimal time and region for the VM allocation. OPA then sends the resulting mutation patches back to the API server which applies them to the VmTemplate CR.

\subsection{Open Policy Agent integration}
\label{opa_integration}


Our system leverages Open Policy Agent (OPA) and the Policy-as-Code paradigm to implement the workload scheduling mechanism in a way that supports user extensibility. 
This is done in two phases.
The OPA component first ensures the compliance of the scheduler component with user-defined policies related to latency and legal constraints (QoS, data residency, etc.) by applying restrictions to scheduling, such as limiting the eligible regions or the available cloud providers. 
Then, it applies a main policy named ``scheduling outcome policy'' which sends real-time requests to invoke the scheduler component with the characteristics of the VM to schedule and the user-defined constraints. The resulting scheduling decisions, i.e. the optimal time and region, are then encoded into Kubernetes resources.
More precisely, the OPA component assumes the role of a Kubernetes Mutating Webhook server that returns the \textit{AdmissionReview} response to the Kubernetes API server. The \textit{AdmissionReview} response contains the decision of the policy evaluation and scheduling outcome as JSON patch operations to be applied to the VmTemplate resource by the Kubernetes API server. 
At the end of this process, the generic VmTemplate resources representing the request for a VM to allocate will be mutated with the optimal cloud provider, time and region in which to be provisioned, as determined by evaluation of user-defined policies and the application of the scheduler. Then, as described in sections \ref{sec:generic-to-specific} and \ref{sec:wait-logic}, once the scheduling time is reached, the mutated VmTemplate resource will then be used to generate the specific CR of a cloud provider leveraging the CDC component.

OPA periodically polls the policy bundles from an external container registry (e.g., DockerHub) to ensure that the policies are up to date.
The system is designed to be highly flexible and extensible, allowing for the addition of new policies and data mappings as needed.

\subsection{Carbon intensity forecaster}
\label{forecaster}

In order to provide the scheduler with information about carbon intensity, a forecaster component was implemented within our architecture. This enables the system to determine carbon intensity measurements in future time periods and at geographically distinct regions to decide the optimal schedule for workloads to reduce their emissions. More specifically, let $\{c_t\}_{t=1}^T$ be a multivariate time series sequence with $c_t$ containing the observed carbon intensity at time $t$, as well as other descriptive variables, and $p$ be the context length i.e. the number of past information used to predict future values. The goal of the forecaster is to learn a function $f$ that maps such sequence to future values for carbon intensity:
$$
\hat{c}_{T+h}=f(c_T, c_{T-1}, ..., c_{T-p+1})
$$
with $h$ being the time horizon i.e. how may hours of carbon intensity will be forecasted. In our case, it is set to 96 as we forecast the next 4 days.

As previously mentioned, the implementation of the forecaster component is based on the 1024 context-length version of the Tiny-
TimeMixers (TTMs) model.
Fine-tuning was performed on the pre-trained model leveraging Electricity Maps' data. More specifically, the moving window technique was applied to create training example tuples as $(X\in\mathbb{R}^{1\times sl}, Y\in\mathbb{R}^{1\times fl})$ where $sl$ is the context length and $fl$ is the forecast length. Considering the time series of length $N=100000$ hourly points, $sl=1024$ and $fl=96$, we generated $N-(sl+fl)=99392$ training examples, with splits of 70\% for training, 20\% for validation and 10\% for testing. From our hyper-parameter benchmarking, the best results were achieved in $10$ epochs using a batch size of $64$, an initial learning rate of $0.00001$ with AdamW optimizer, and a dropout rate of $0.3$. Fine-tuning was then performed to obtain one model for each region. 

\begin{figure}[htb]
  \centering
  \includegraphics[width=1\linewidth]{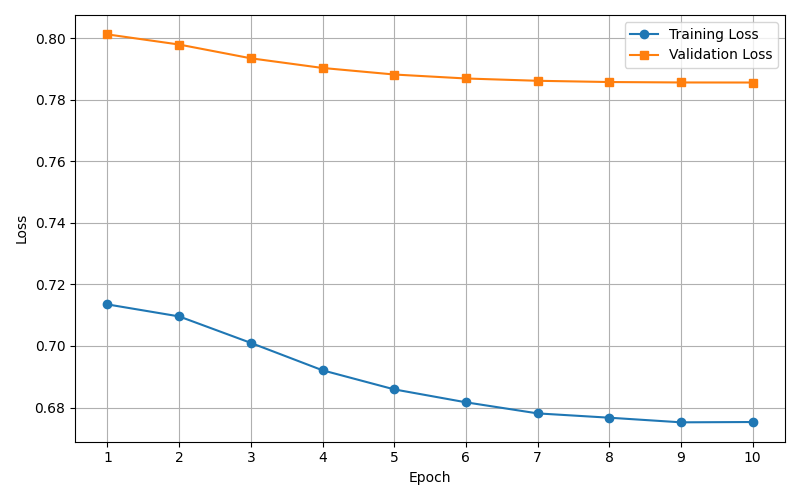}
  \caption{Fine-tuning of the forecaster model on the US-TEX-ERCO region.}
  \label{fig:finetuning}
\end{figure}

Figure \ref{fig:finetuning} shows an example of the fine-tuning process as performed on the US-TEX-ERCO region.

The fine-tuned models are then deployed within our architecture leveraging KServe, as described in the next section. Furthermore, an hourly cron job is configured to create inference requests to forecast the next $n$ days of carbon intensity at each region, with $n$ set to 4 days by default. The resulting forecasts are then pushed by the forecaster component into a CrateDB database \footnote{https://cratedb.com/} with an SQL query.

\subsection{MLOps Kubernetes infrastructure}

\begin{figure}[htb]
  \centering
  \includegraphics[width=0.8\linewidth]{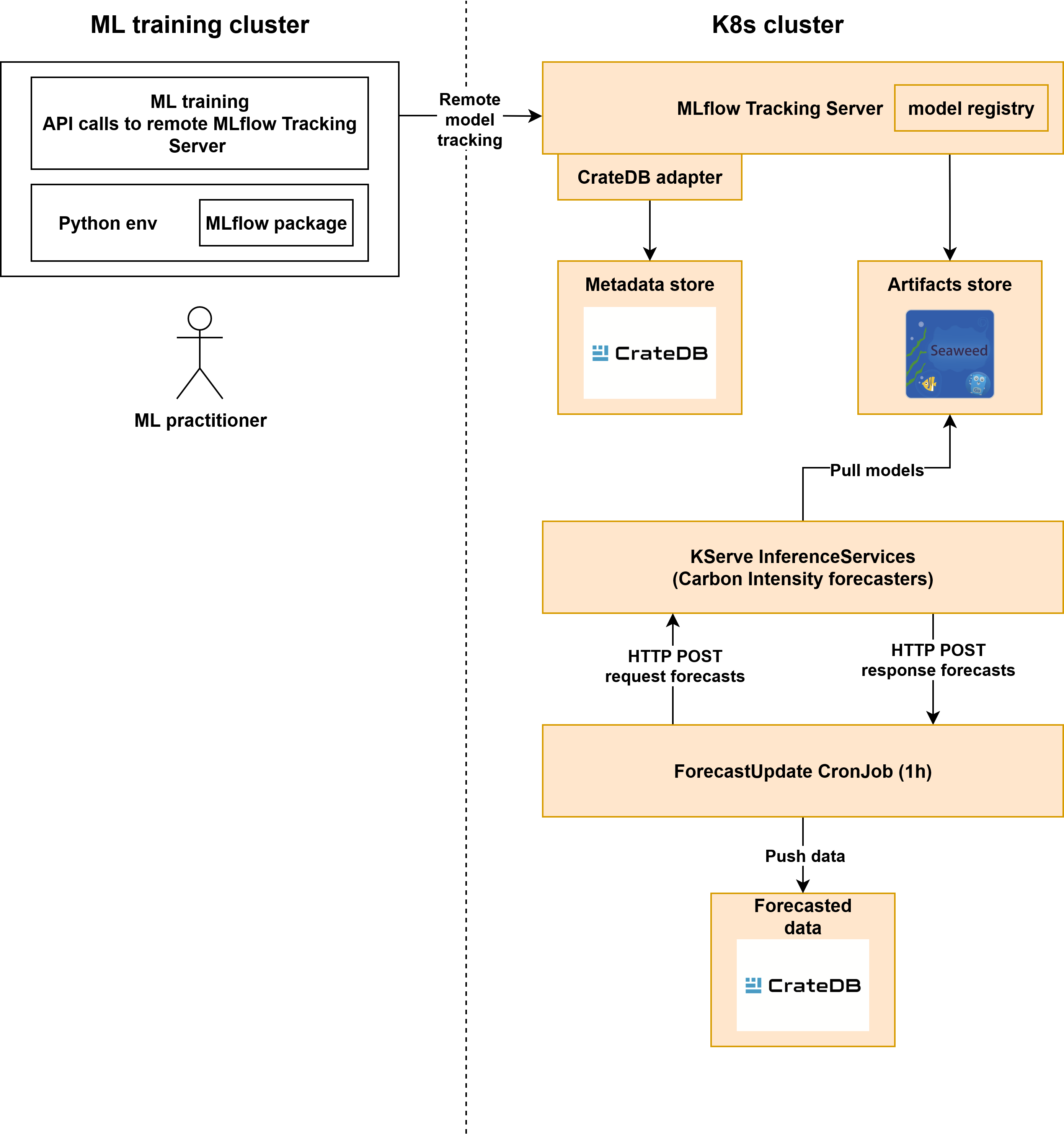}
  \caption{MLOps configuration within the proposed system, with MLFlow for model tracking and KServe for runtime inference.}
  \label{fig:mlflow_config}
\end{figure}

To support the MLOps workflow of the forecaster models finetuning and deployment, MLflow \footnote{https://mlflow.org/} is deployed within the system. This is done using a loosely coupled architecture as can be seen in figure \ref{fig:mlflow_config}, where the MLflow Tracking Server is decoupled from its storage. This configuration is the most common in production environments, as it allows for better scalability and environment flexibility. More specifically, the MLflow storage is composed of a metadata store (backend store) and an artifacts store. The metadata store chosen for the system is CrateDB, while the artifact store is the SeaweedFS object storage service \footnote{https://github.com/seaweedfs/seaweedfs}.

For the deployment of the finetuned models, KServe InferenceServices \footnote{https://github.com/kserve/kserve} are employed within our system. More specifically, one InferenceService instance is deployed for each model, each corresponding to a different region. Another InferenceService is deployed for a generic model that is used as a fallback if a specified region is not available.
This setting introduces quite a large amount of overhead in terms of resources, since each InferenceService sets up a whole new set of resources (e.g., large Kubernetes pods with underlying model serving environments) for each model. It is an acceptable trade-off since the best carbon intensity forecast performances were observed using a dedicated model for each region as described above.

\begin{figure}[htb]
  \centering
  \includegraphics[width=0.9\linewidth]{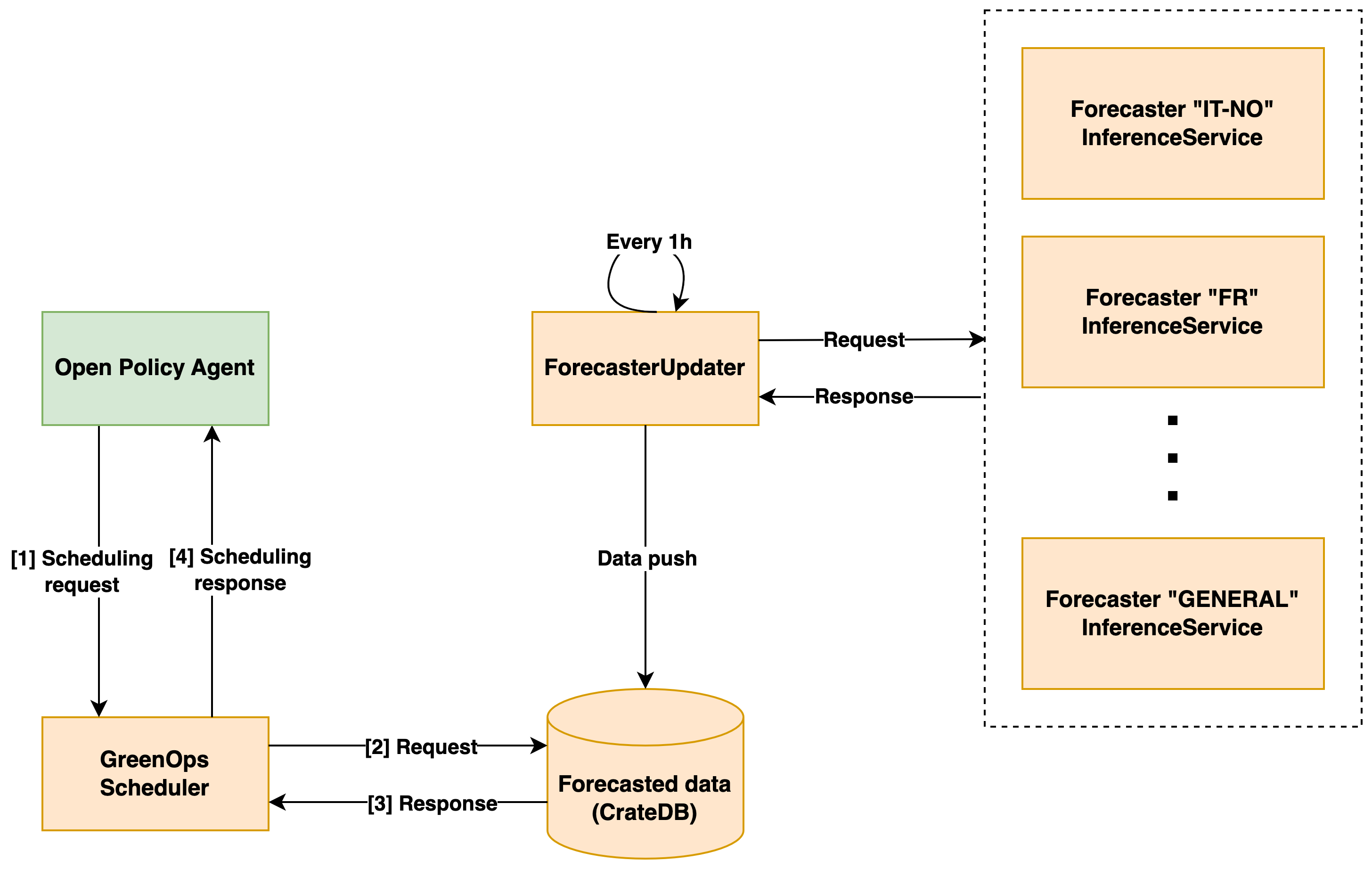}
  \caption{Interactions between the scheduler and forecaster components: periodic push of forecasted data by the CronJob and retrieval of latest forecasts during scheduling}
  \label{fig:scheduler_forecaster}
\end{figure}

\subsection{Scheduler}

The scheduler component consists of a lightweight HTTP server implemented using the Flask library \footnote{https://flask.palletsprojects.com/en/stable/} to respond to VM allocation requests. They include the tuple defining the characteristics of the VM to schedule as well as the list of eligible regions as determined by policy evaluation. The scheduler then applies the time- and space-shifting scheduling optimization logic implemented with the PuLP library. To this end, the latest carbon intensity forecasts are pulled from the CrateDB database. As the output of the scheduling, the optimal time and region are returned in the response to the OPA server.

\subsection{System limitations}

A limitation of our general approach is that only resources supported by the cloud provider’s Kubernetes operator can be provisioned in a seamless way. Indeed, while most cloud resources available in a cloud provider’s portfolio are available as Kubernetes CRs, this is not guaranteed.
This fact poses few additional considerations other than limited resource availability:
\begin{itemize}
    \item Dependence on operator updates: cloud providers may arbitrarily extend or modify the set of resources supported by their Kubernetes operators over time. This is mitigated by the provider's update policies and deprecation notices of API versions.
    \item Vendor-specific implementations: for the same class of resources (e.g., virtual machines), the structure and fields of the CRs may vary in a considerable way between cloud providers. This issue is mitigated through the routing guards in the Helm templates.
\end{itemize}

Despite these constraints, the system architecture remains highly adaptable thanks to the provided mitigation strategies. Nonetheless, future enhancements could incorporate additional or alternative provisioning mechanisms. 
An example of an alternative implementation could be the direct API interactions with cloud providers to bypass operator limitations. In fact, cloud provider operators typically leverage these APIs under the hood to interact with the cloud provider’s services.
Another approach could involve the development of custom operators or controllers to manage specific resource types not supported by existing operators.

\section{Evaluation}
\label{evaluation}

We conducted an experimental evaluation of our proposed system by comparing it to a baseline carbon-agnostic round-robin scheduler to quantify the reduction in carbon footprint achieved by our carbon-aware approach. The round-robin schedule was selected as a baseline for comparison, as it is a reasonable approach to resource allocation that is fair but does not consider the carbon footprint of its decisions. 

\subsection{Evaluation setup}

The evaluation was conducted using CloudSim Plus \cite{silva2017cloudsim}, a state-of-the-art simulator for cloud computing infrastructures and application services. In particular, we modeled a multi-cluster scenario with data centers located at Azure cloud regions. Each cluster hosts 200 simulated physical machines modeled with the specifications of Dell PowerEdge XR8620T rack servers (equipped with Intel Xeon Gold 6433N 32 cores @ 2.00GHz CPU and 256GB RAM). The power consumption profile of each machine is modeled with data from SPECpower \cite{spechost_website}, the industry standard benchmark for performance and energy measurements. In this way, we were able to quantify the carbon footprint of the executed workloads according to the power consumption of the host machines and the carbon intensity of the region in which the data center is located as:
\[
\text{Emissions} = \sum_t^T \text{Power}(t) * \text{CI}(t)
\]
where \text{Power}(t) is the power consumption (in kWh) of host machines at hour $t$ and $\text{CI}(t)$ is the carbon intensity (in gCO\textsubscript{2}eq/kWh) of the underlying grid, summed over the hourly timestamps of the evaluation period.

We used historical data on carbon intensity from Electricity Maps to quantify the emissions of the scheduling approaches. Our experimental evaluation spanned from mid-May to mid-June 2022, a period chosen to align with the month-long dataset of VM traces. This timeframe also captures seasonal variations in carbon intensity, such as the increased solar power generation observed during the transition from spring to summer. For the realistic scenario, we used our forecaster component to generate carbon intensity forecasts to cover the same period, enabling a consistent and comparative analysis across both evaluation modes.


\begin{figure*}[h!]
    \centering
    \begin{subfigure}[b]{0.325\linewidth}
        \centering
        \includegraphics[width=\linewidth]{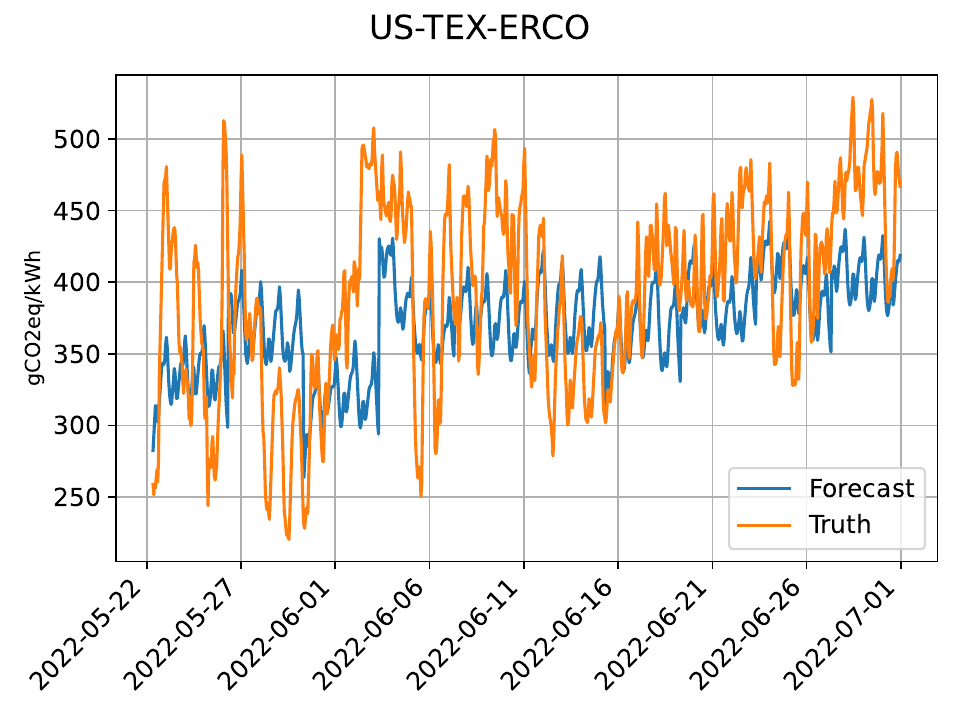}
        \label{fig:forecast-us}
    \end{subfigure}
    \hfill
    \begin{subfigure}[b]{0.325\linewidth}
        \centering
        \includegraphics[width=\linewidth]{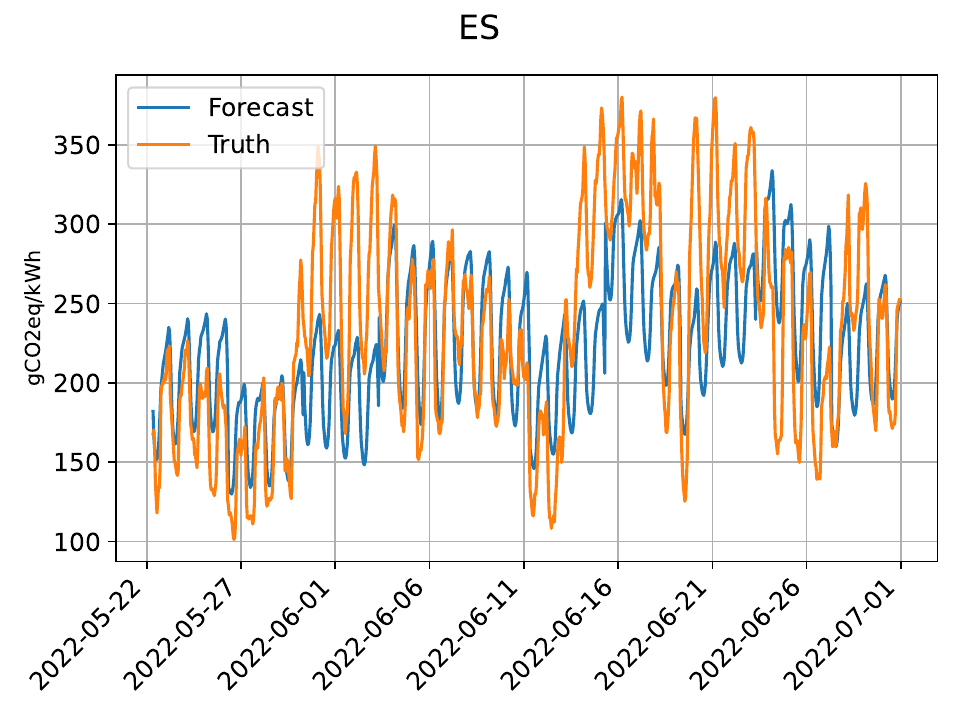}
        \label{fig:forecast-es}
    \end{subfigure}
    \hfill
    \begin{subfigure}[b]{0.325\linewidth}
        \centering
        \includegraphics[width=\linewidth]{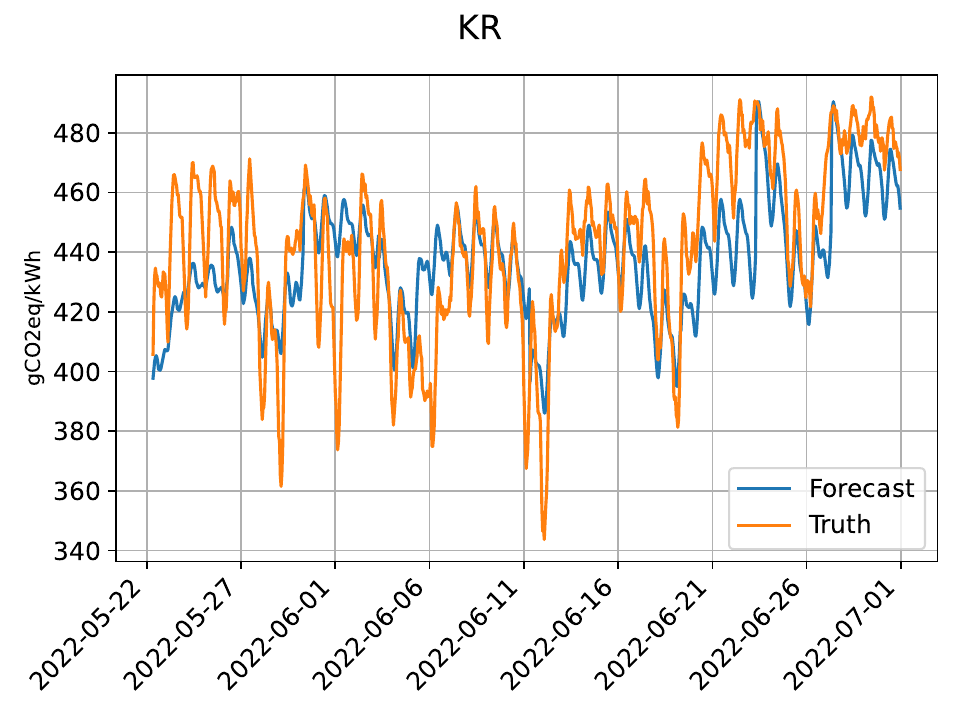}
        \label{fig:forecast-kr}
    \end{subfigure}
    \caption{Examples of carbon intensity forecasts in the testing period for the US-TEX ERCO, Spain and Korea regions.}
    \label{fig:delay-distribution-deadline}
\end{figure*}

\begin{figure*}[]
    \centering
    \begin{subfigure}[b]{0.48\textwidth}
        \centering
        \includegraphics[width=\linewidth]{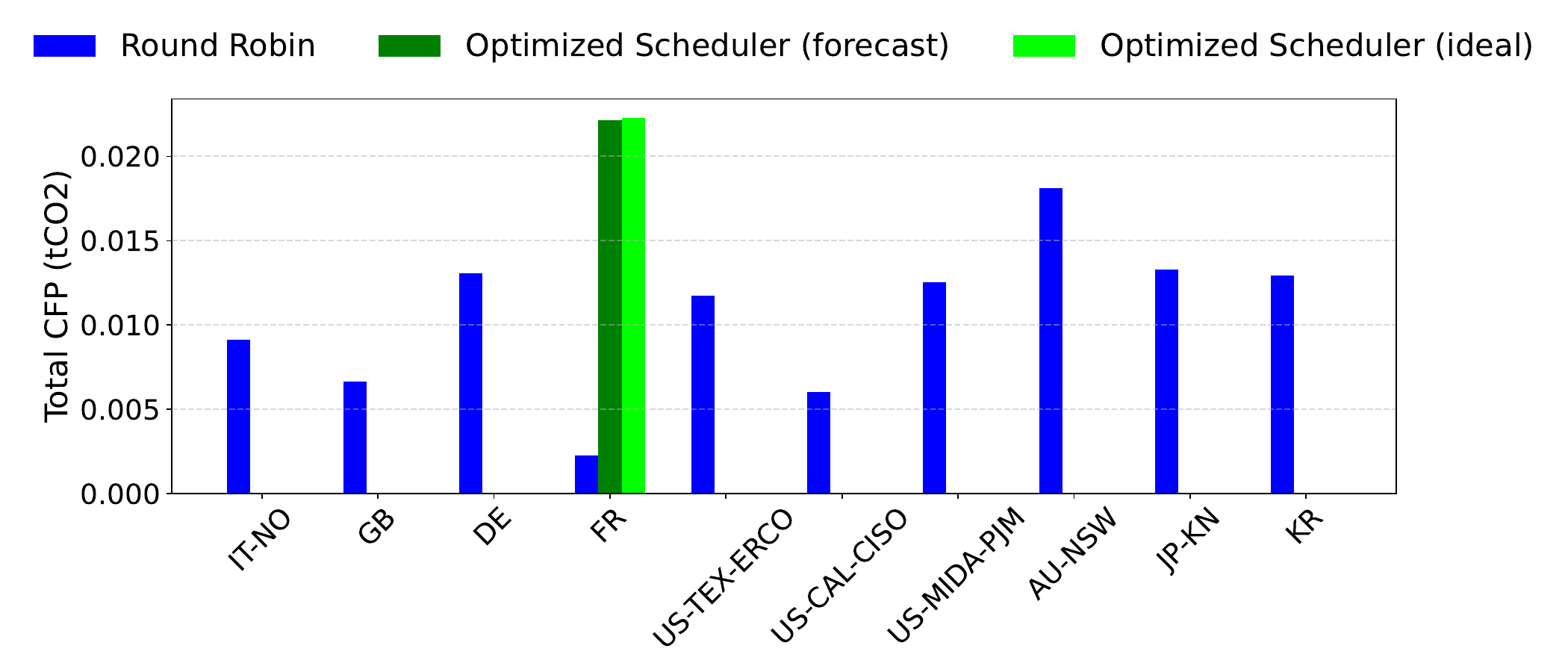}
        \caption{Subset policy, max 50 job per region}
        \label{fig:cfp-subset-50}
    \end{subfigure}
    \hfill
    \begin{subfigure}[b]{0.48\textwidth}
        \centering
        \includegraphics[width=\linewidth]{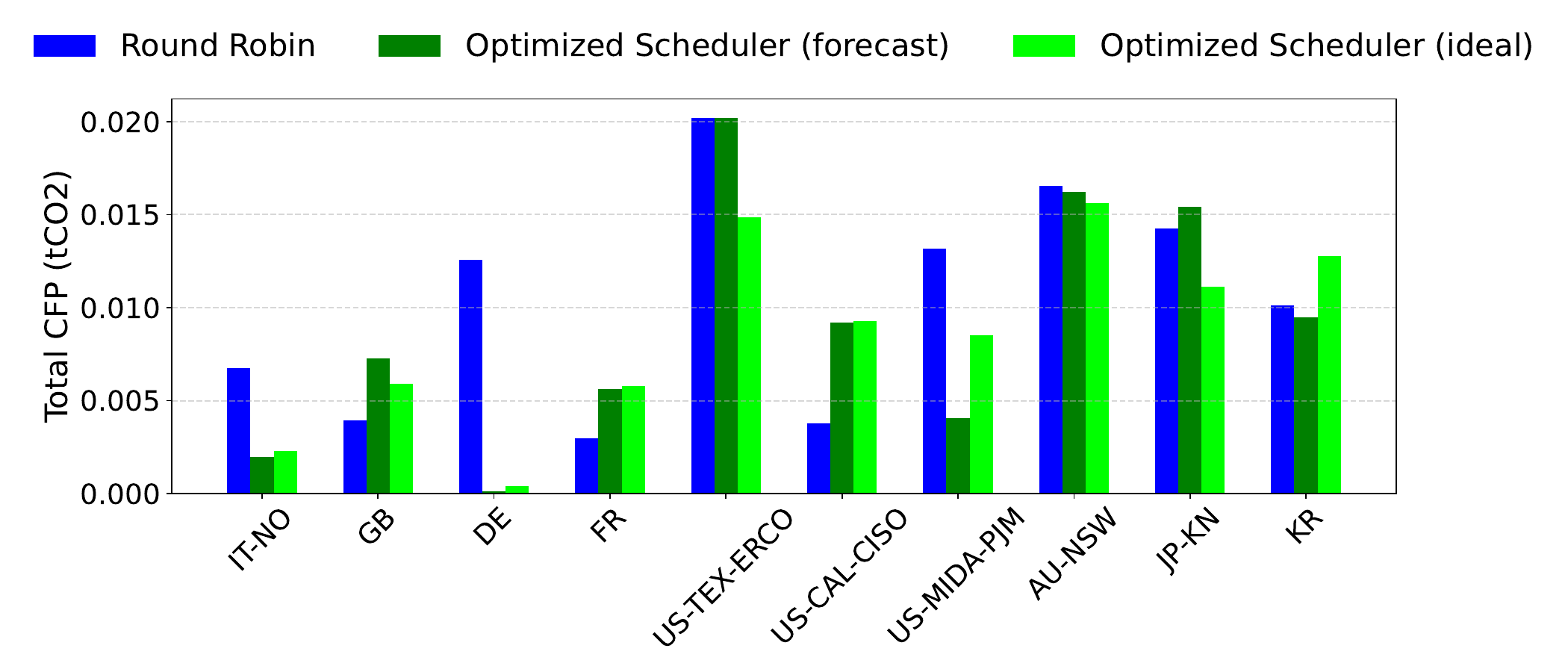}
        \caption{Latency policy, max 5 job per region}
        \label{fig:cfp-latency-5}
    \end{subfigure}
    \caption{Carbon emissions at regions in best- and worst-case scenarios. Strict constraints, as in the latency policy with 5 maximum simultaneous jobs per region, result in more geographically distributed scheduling with higher emissions, while more permissive constraints, as in the subset policy with 50 maximum simultaneous jobs per region, allow for greater emissions reduction by scheduling more jobs to the optimal region.}
    \label{fig:cfp-best-worst-case}
\end{figure*}

To ensure realism in the VM allocation use cases, we used VM request traces from the publicly available Azure dataset \cite{cortez2017resource}. It comprises over two million VM requests from first-party Azure workloads over the span of one month. Each trace includes timestamps for VM allocation and deallocation, as well as resource requirements such as the number of CPU cores and RAM. For our evaluation, we selected 1,000 batches each consisting of 1,000 VM allocation requests. We sampled them from the dataset considering a VM lifetime duration between 6 and 24 hours. We considered ranging values of 6, 12, 24, and 48 hours of job completion deadline (accounting for job duration) to evaluate the benefits of the time-shifting scheduling supported by our system.

We also evaluated a range of values for the maximum amount of jobs allowed to run simultaneously at each region. This analysis allowed us to assess constraint (3) and its impact on the "black hole" problem. Based on our VM requests batch size, we tested the scheduling approaches with limits of 5, 10, 20, and 50 maximum simultaneous jobs per region. 
As expected, with a lower number of allowed simultaneous jobs the optimized scheduler tends to behave similarly to the baseline round-robin scheduler. In this case, VMs cannot always be allocated to the best region but are instead distributed among other sub-optimal regions due to resource constraints. With a higher number of allowed jobs instead, they all tend to be scheduled to the region with the lowest carbon intensity, leading to greater reductions in carbon footprint but also the problem of the black hole.

To determine the regions in which we performed our evaluation, we leveraged the user policy support provided by our system. We evaluated three scenarios using representative policies to determine eligible regions for VM allocation. Building on the Azure traces dataset, we constrained the analysis to regions available within the Azure cloud infrastructure. The evaluated policies are as follows:
\begin{enumerate}
    \item \textit{Subset}: a subset of representative regions we selected to evaluate the scheduling decisions. Based on the available Electricity Maps 
    data, we considered regions with ranging mean and variability of Carbon Intensity while covering most continents. Namely, this policy restricts the eligible regions to North Italy, Great Britain, Germany, France, US (Ercot, California ISO, PJM) Australia (New South Wales), Korea, and Japan (Kansai). 
    \item \textit{GDPR}: a privacy-driven policy limiting eligible regions to those within the European Union to enforce compliance with the data residency constraint of the GDPR. It restricts eligible regions to Ireland, Central Sweden, Netherlands, France, Germany, North Italy, Poland, and Spain.
    \item \textit{Latency}: a performance-driven policy that allows only regions with a maximum network latency of 50 milliseconds or less from the origin location of the user. It considers the regions from the subset policy and assigns a random origin location to the requests.
\end{enumerate}

The three scenarios defined by the application of the policies were evaluated against both the baseline, carbon-agnostic round-robin scheduler, and the carbon-aware optimized scheduler presented in our architecture. Furthermore, the application of the scheduler was evaluated in both an ideal and a realistic scenario. In the first case, we assume perfect knowledge of the future carbon intensity using historical data, while for the latter the future carbon intensity is determined by the application of the forecaster component of our architecture. This allowed us to quantify a theoretical upper bound on the achievable reduction in carbon footprint.


\subsection{Evaluation results}

\begin{figure*}[h!]
    \centering

    \begin{minipage}{\textwidth}
        \centering
        \begin{subfigure}[b]{0.48\linewidth}
            \includegraphics[width=\linewidth]{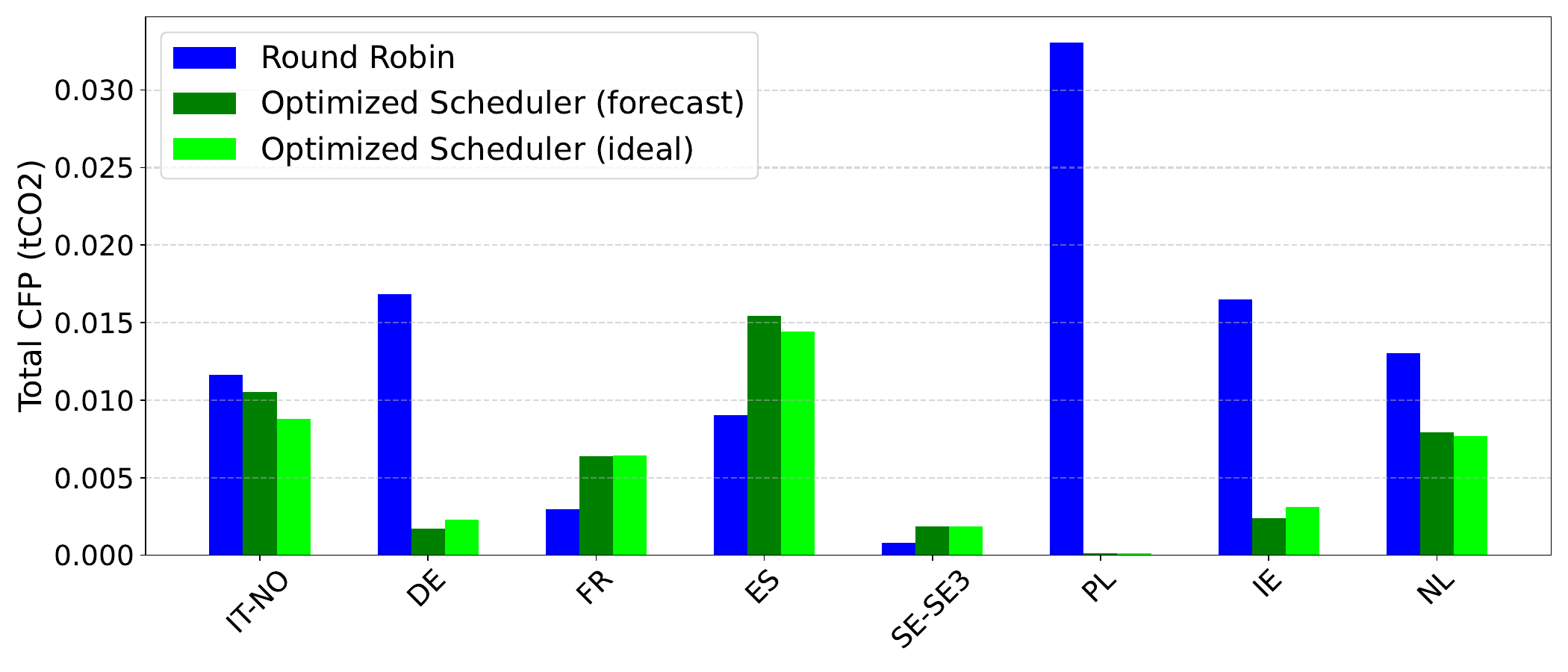}
            \caption{GDPR policy, 5 max jobs: carbon footprint}
        \end{subfigure}
        \hfill
        \begin{subfigure}[b]{0.48\linewidth}
            \includegraphics[width=\linewidth]{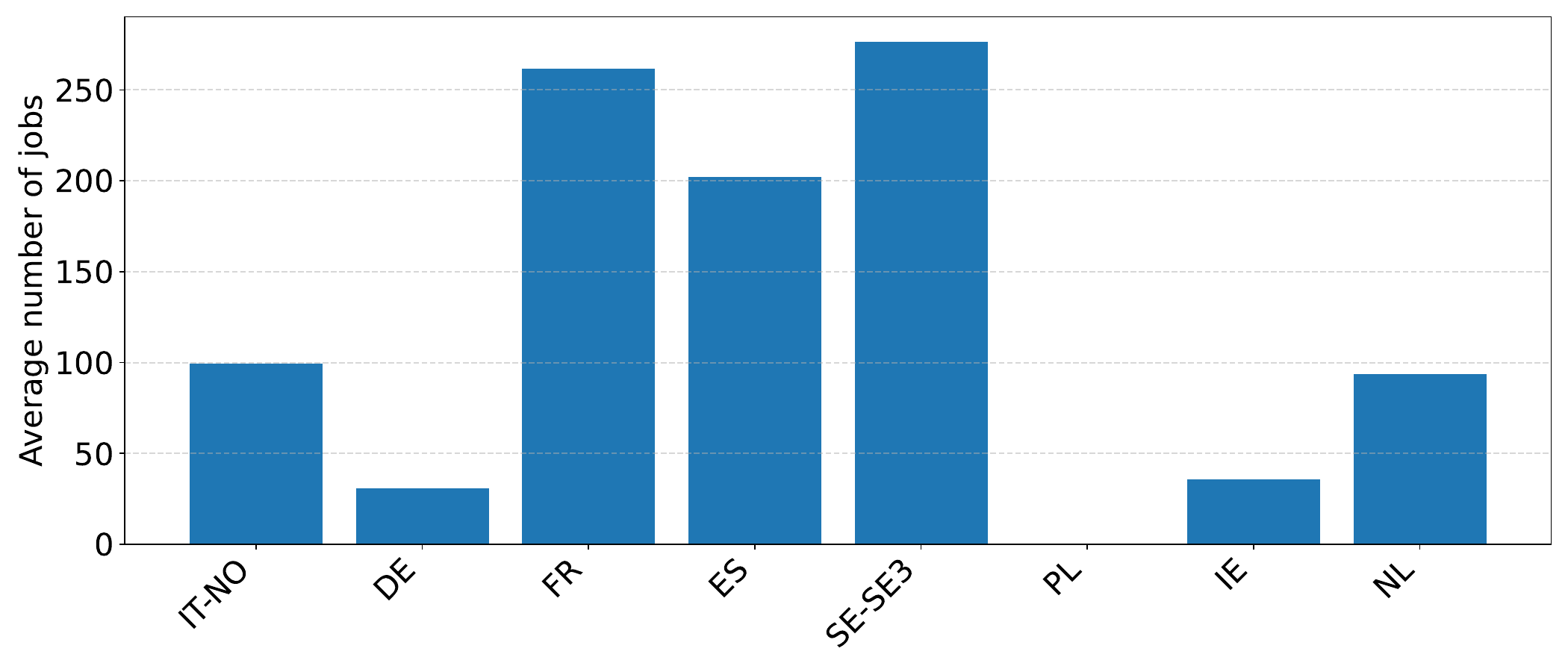}
            \caption{GDPR policy, 5 max jobs: job distribution}
        \end{subfigure}
    \end{minipage}


    \begin{minipage}{\textwidth}
        \centering
        \begin{subfigure}[b]{0.48\linewidth}
            \includegraphics[width=\linewidth]{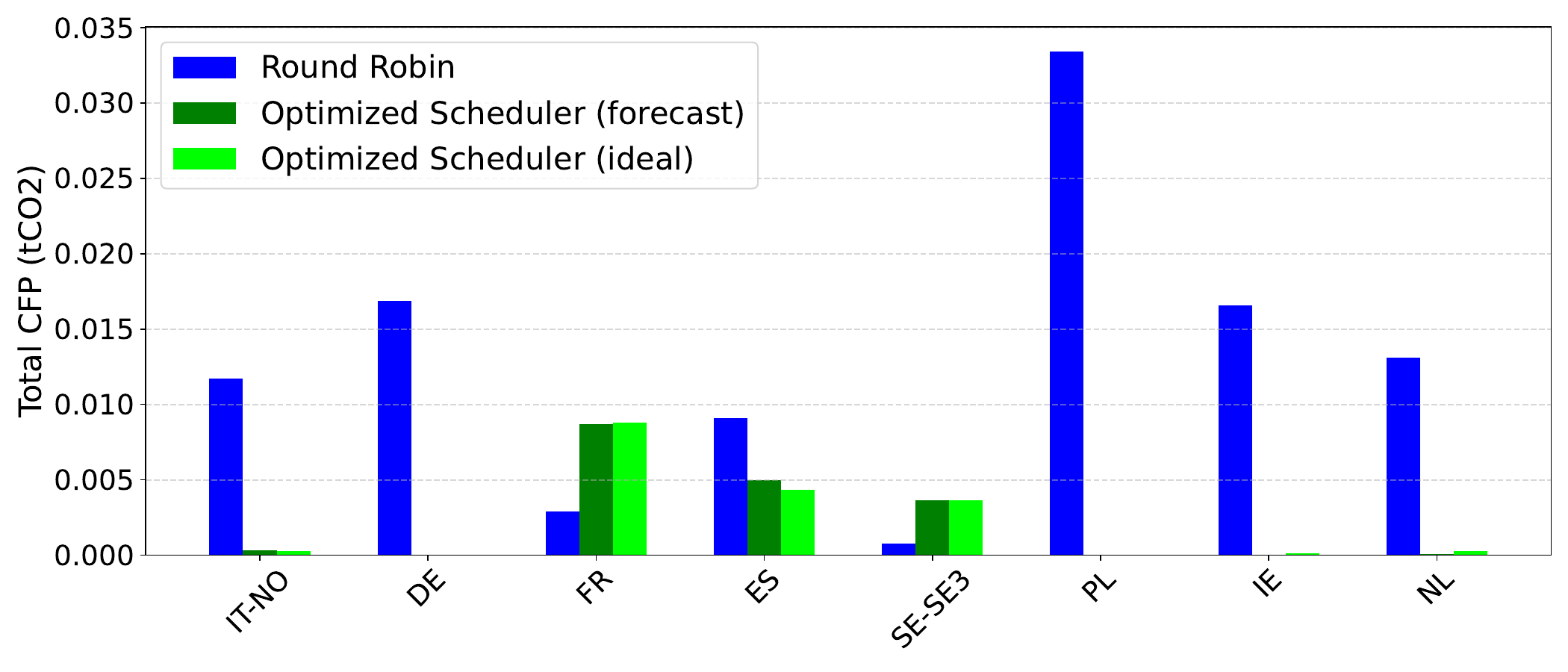}
            \caption{GDPR policy, 10 max jobs: carbon footprint}
        \end{subfigure}
        \hfill
        \begin{subfigure}[b]{0.48\linewidth}
            \includegraphics[width=\linewidth]{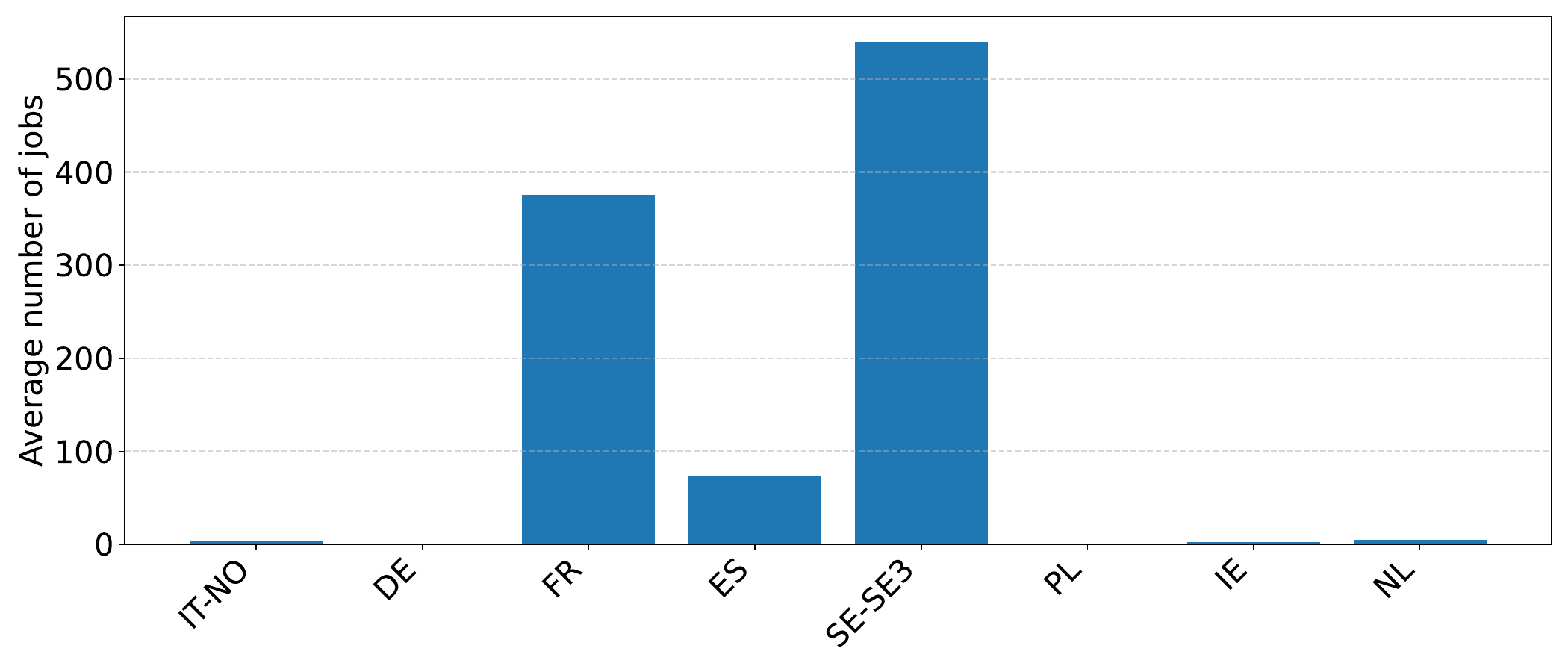}
            \caption{GDPR policy, 10 max jobs: job distribution}
        \end{subfigure}
    \end{minipage}


    \begin{minipage}{\textwidth}
        \centering
        \begin{subfigure}[b]{0.48\linewidth}
            \includegraphics[width=\linewidth]{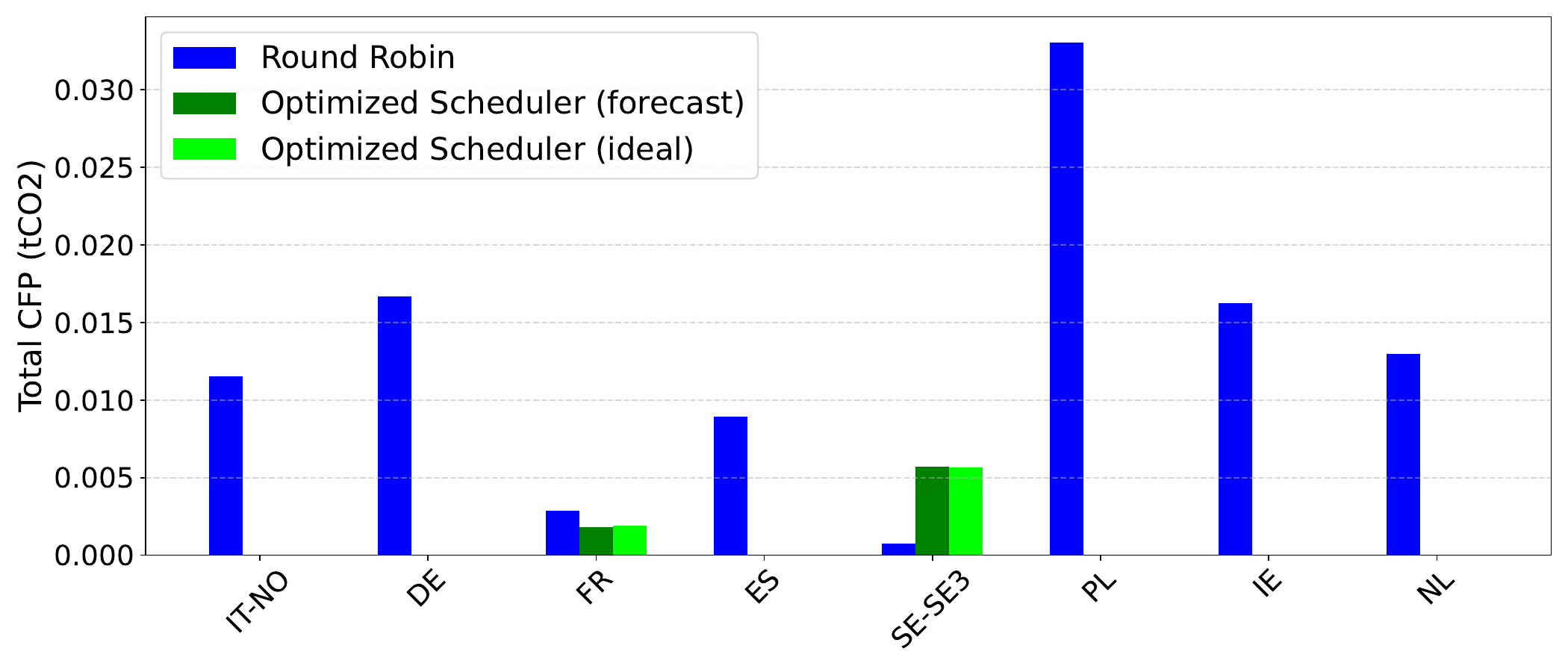}
        \caption{GDPR policy, 20 max jobs: carbon footprint}
        \end{subfigure}
        \hfill
        \begin{subfigure}[b]{0.48\linewidth}
            \includegraphics[width=\linewidth]{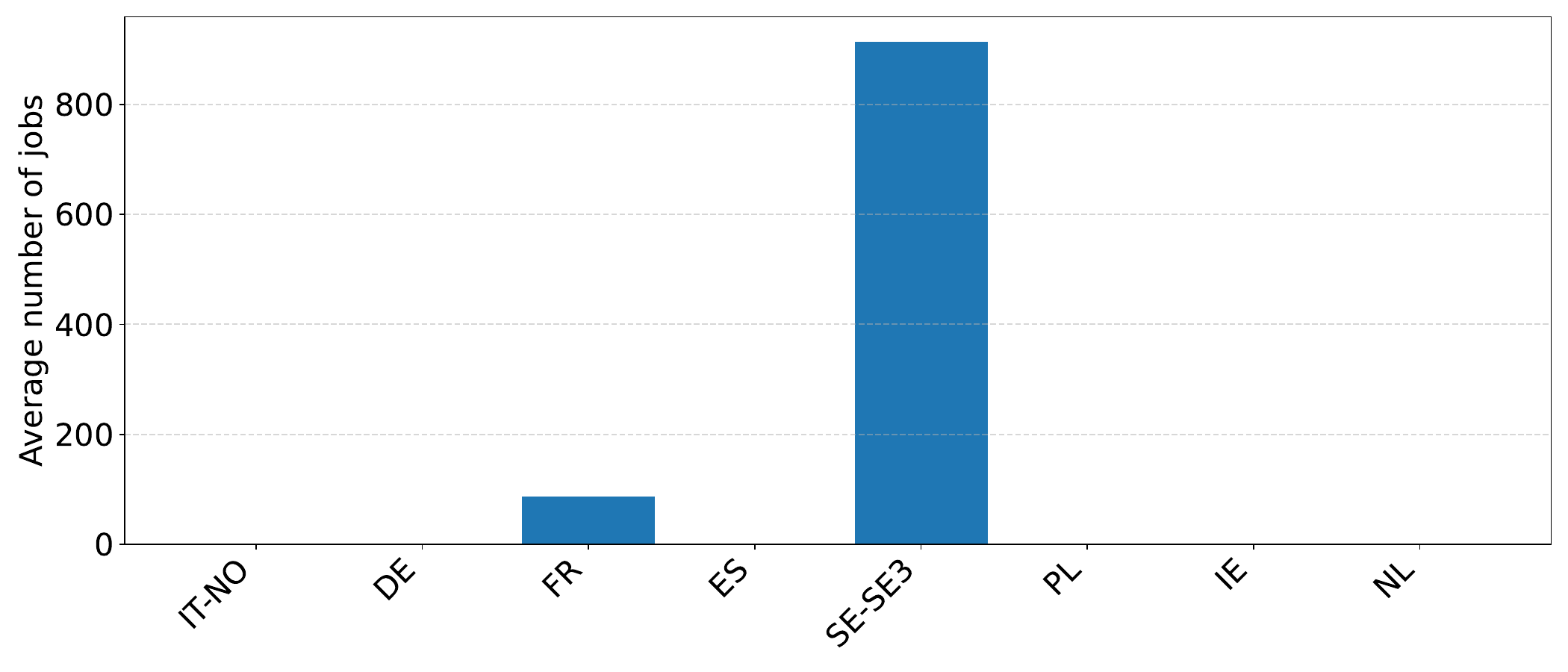}
            \caption{GDPR policy, 20 max jobs: job distribution}
        \end{subfigure}
    \end{minipage}


    \begin{minipage}{\textwidth}
        \centering
        \begin{subfigure}[b]{0.48\linewidth}
            \includegraphics[width=\linewidth]{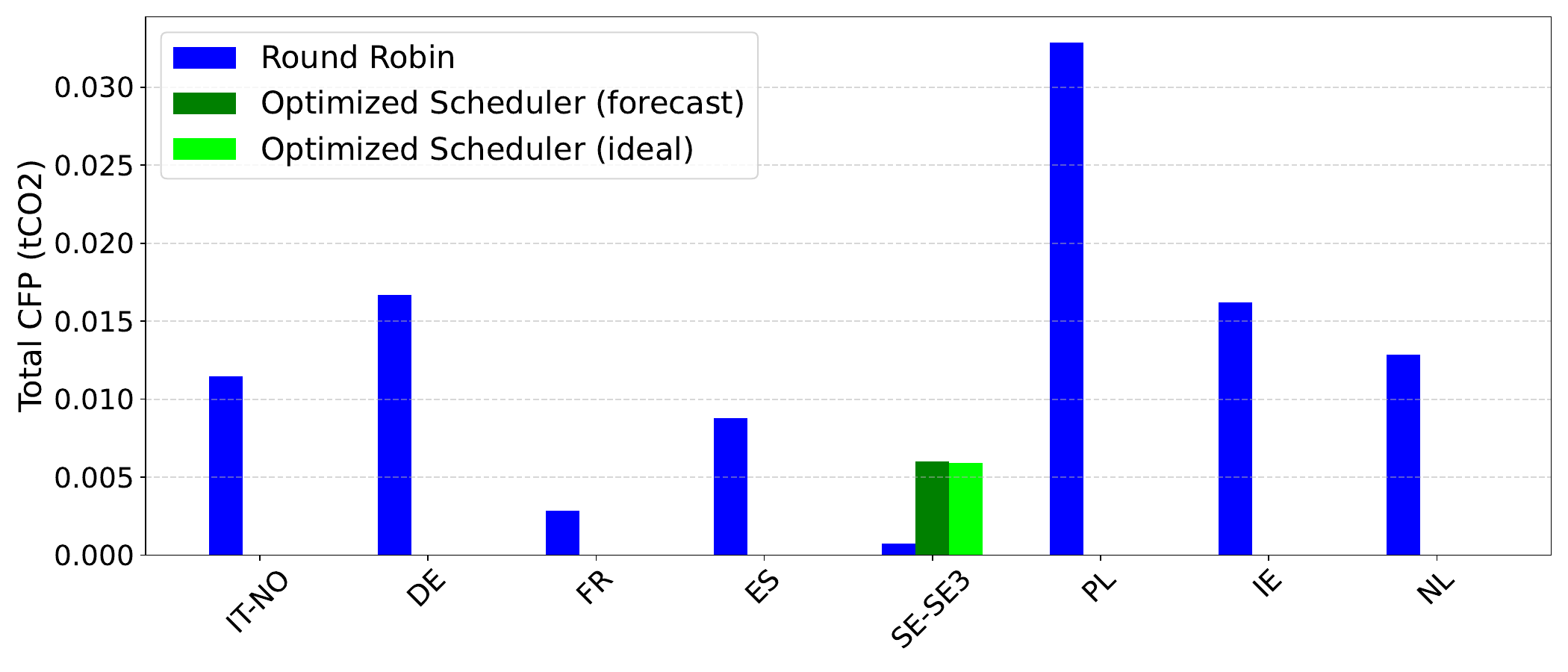}
            \caption{GDPR policy, 50 max jobs: carbon footprint}
        \end{subfigure}
        \hfill
        \begin{subfigure}[b]{0.48\linewidth}
            \includegraphics[width=\linewidth]{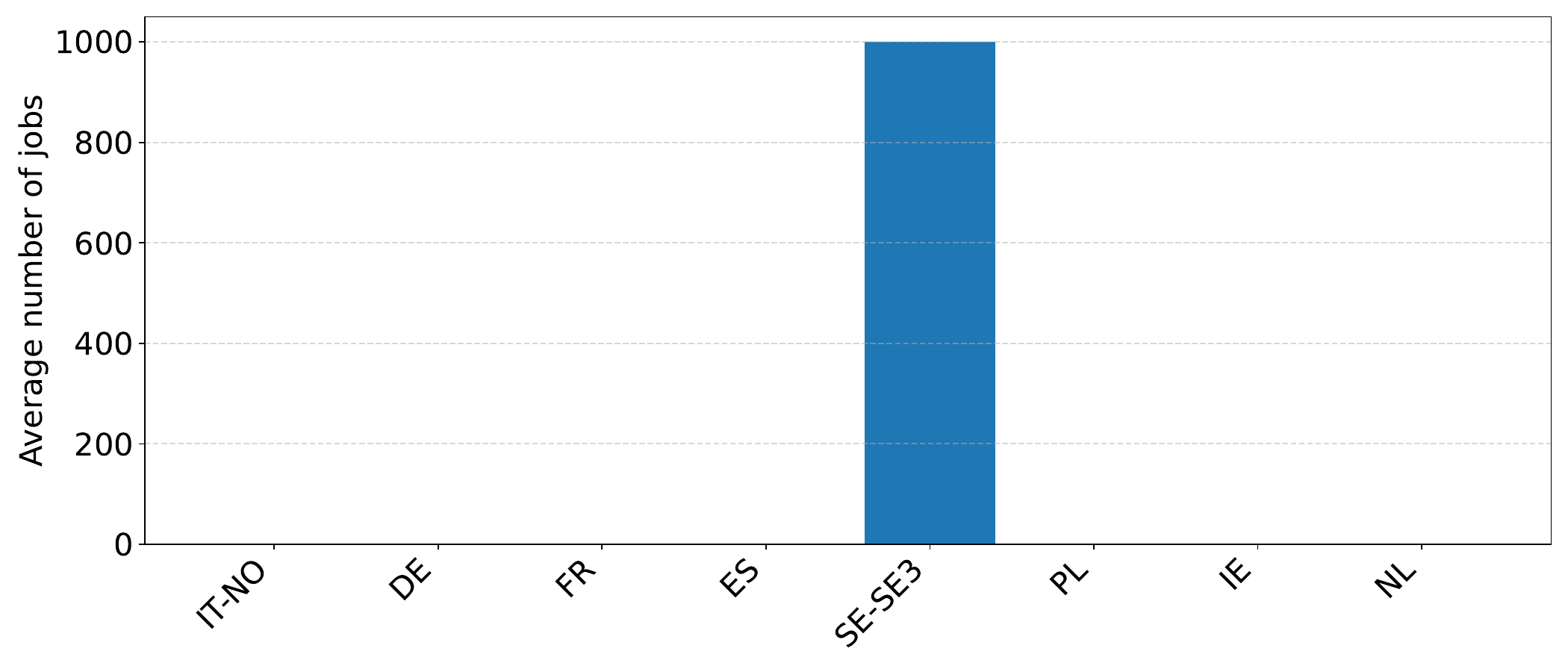}
            \caption{GDPR policy, 50 max jobs: job distribution}
        \end{subfigure}
    \end{minipage}

    \caption{Jobs carbon footprint and distribution across regions for different GDPR policy values.}
    \label{fig:black-hole}
\end{figure*}


Simulation results demonstrate that our proposed system effectively applied both temporal and spatial schedule shifting to achieve significant reductions in the carbon footprint (CFP) of cloud workloads (Figure \ref{fig:cfp-best-worst-case}). 

All evaluated policies and parameter configurations showed improvements, including the more restrictive scenarios under strict limitations on the availability of regions and resources. In the best-case scenario with the most permissive configuration -- using the subset policy and allowing up to 50 simultaneous jobs per region -- the system achieved a CFP reduction of 79.25\% compared to the baseline carbon-agnostic round-robin scheduler. This result holds in both ideal scheduling with historical data and in the realistic scheduling with the forecaster application. In the more restrictive scenario -- using the latency policy and limiting regions to a maximum of 5 simultaneous jobs -- we instead observed a 13.46\% reduction with forecasted data and a 16.35\% reduction with perfect knowledge.

Another key outcome of our simulations is the relatively small difference in carbon footprint reductions achieved when using perfect foresight about future carbon intensity compared to those achieved under forecasted data. This results highlights the effectiveness of our forecaster component, which enables our system to compute schedules that closely approximate the theoretical optimum.

Evaluating ranging values of the maximum number of allowed simultaneous jobs running at each region reveals the effects of the so-called "black hole" problem, where all jobs are scheduled to the region with the lowest carbon intensity, even when such scheduling is ultimately unfeasible. For example, as can be seen in Figure \ref{fig:black-hole}, jobs scheduled under the GDPR policy end up being allocated to a single region when the limit of allowed simultaneous jobs per region becomes more permissive. In contrast, a stricter limit promotes a more balanced distribution of jobs across regions. This difference in allocation directly affects carbon savings: a limit of 5 maximum simultaneous jobs per region yields a nearly 50\% reduction in carbon emissions when compared to the carbon-agnostic baseline, while a more permissive limit of 50 jobs per region achieves up to a 90\% reduction by scheduling nearly all jobs to lowest-carbon region. 
However, such an ideal allocation may not be feasible in practice. This is because data center capacity constraints and provider-level resource management policies can prevent all jobs from being scheduled to a single region, particularly under high demand.
Moreover, as already mentioned, increases in electricity demand resulting from increased computing services might require the activation of peaking power generators, which typically rely on fossil fuels. This, in turn, may raise the levels of carbon intensity thus limiting the actual reductions in carbon footprint. 
As such, accurately estimating the limit on the allowed simultaneous jobs per region is of critical importance to determine both effective and practical scheduling outcomes.

\begin{figure*}[b]
    \centering
    \begin{subfigure}[]{0.24\linewidth}
        \centering
        \includegraphics[width=\linewidth]{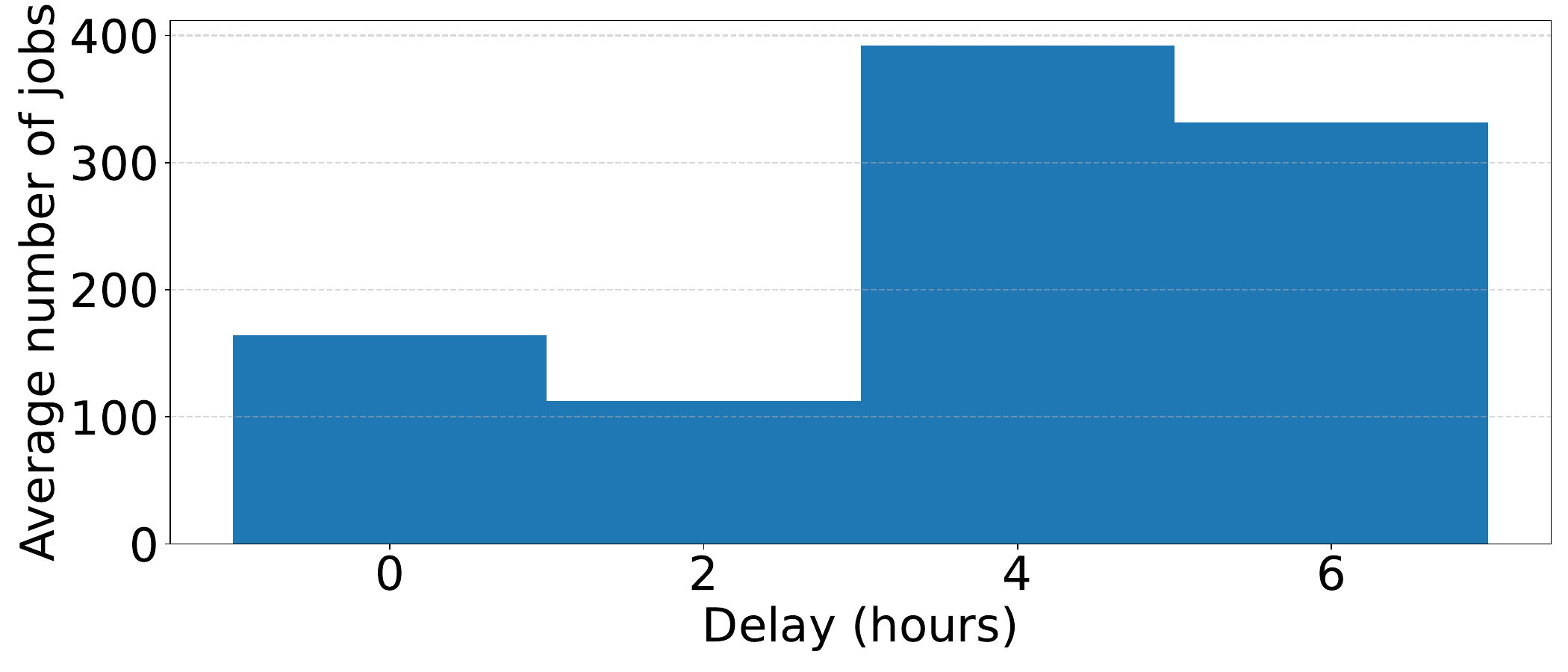}
        \caption{6 hours}
        \label{fig:delay-6}
    \end{subfigure}
    \hfill
    \begin{subfigure}[]{0.24\linewidth}
        \centering
        \includegraphics[width=\linewidth]{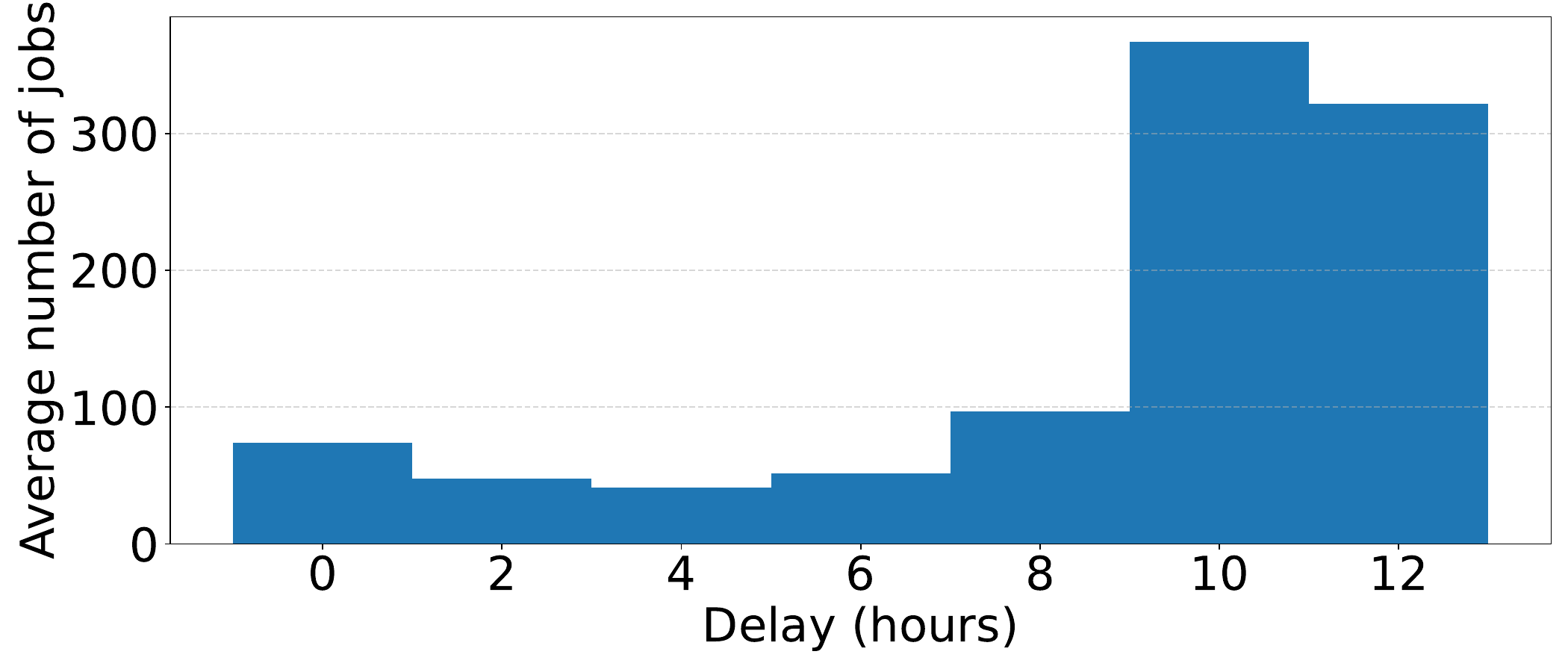}
        \caption{12 hours}
        \label{fig:delay-12}
    \end{subfigure}
    \hfill
    \begin{subfigure}[]{0.24\linewidth}
        \centering
        \includegraphics[width=\linewidth]{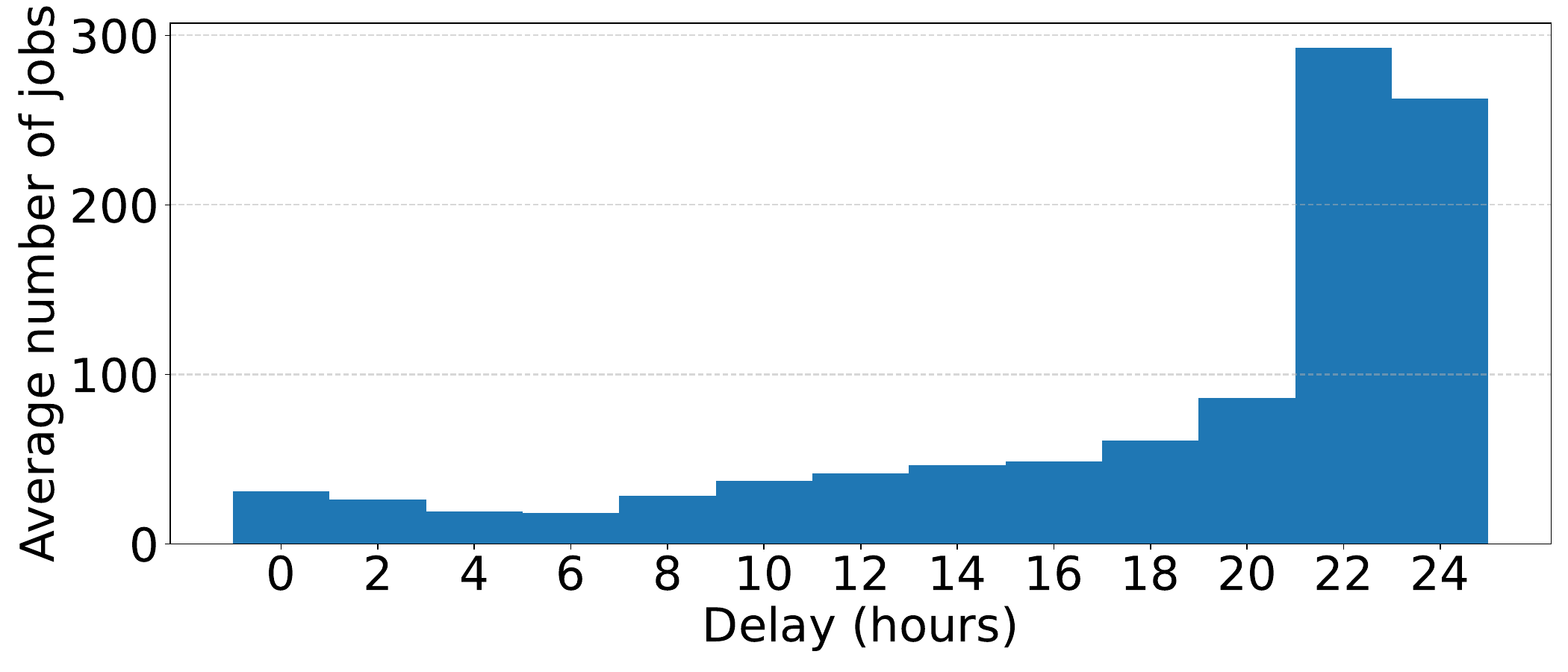}
        \caption{24 hours}
        \label{fig:delay-24}
    \end{subfigure}
    \hfill
    \begin{subfigure}[]{0.24\linewidth}
        \centering
        \includegraphics[width=\linewidth]{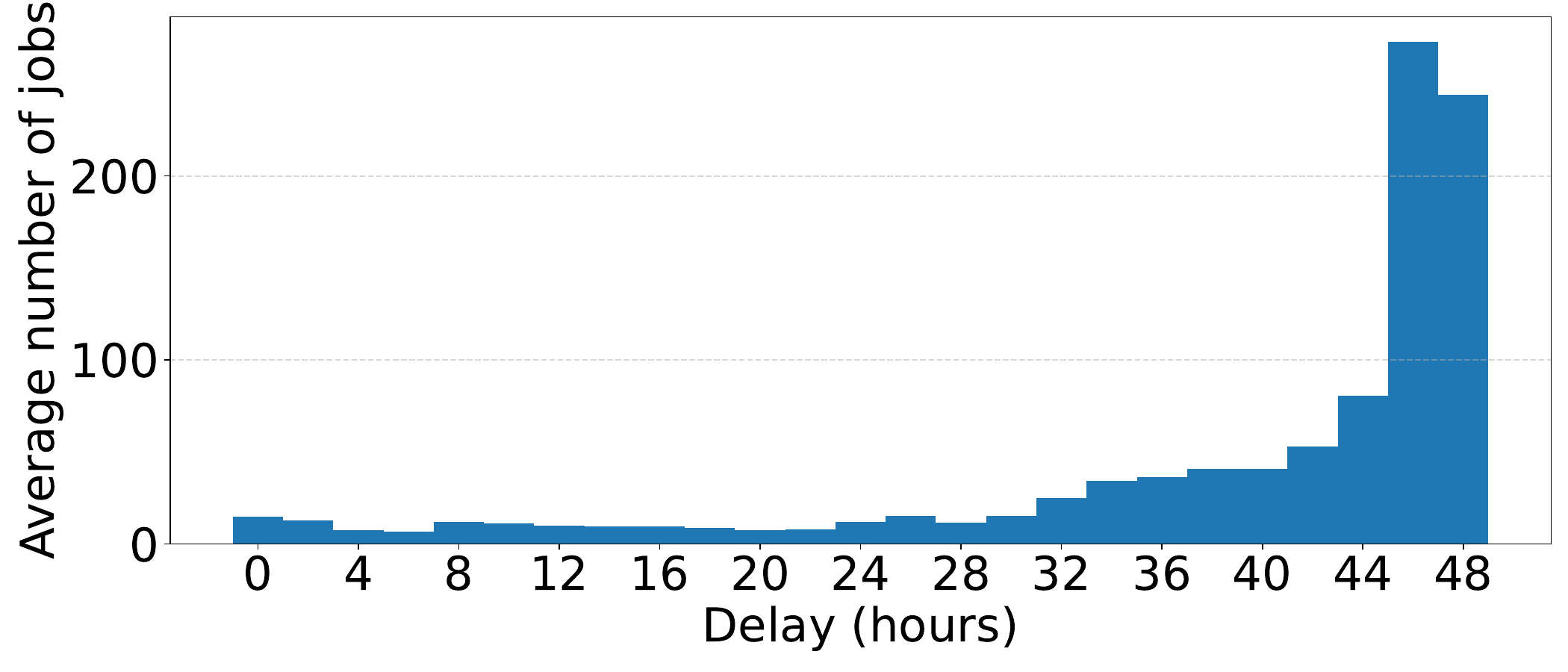}
        \caption{48 hours}
        \label{fig:delay-48}
    \end{subfigure}
    \caption{Average delay of job execution at ranging values of deadline margin (subset policy). Increasing the deadline limit of jobs increases the exploitation of time-shifting scheduling mechanisms with greater deferral of job execution.}
    \label{fig:delay-distribution-deadline}
\end{figure*}

Evaluating ranging values for the deadline on jobs execution allowed us to quantify the effects the time-shifting scheduling approach. Simulation results indicates that a larger margin enables greater deferral of job execution, which in turn allows the scheduler to better exploit temporal variations in carbon intensity. This ultimately leads to a greater reduction in carbon footprint. Specifically, under the subset policy allowing a maximum of 10 simultaneous jobs per region, the system achieves up to a 13\% reduction in emissions when the deadline margin is extended to 48 hours ($0.036$ tCO\textsubscript{2} total emissions), compared to a more constrained margin of 6 hours ($0.041$ tCO\textsubscript{2} total emissions).


Moreover, the deadline flexibility also influences the spatial distribution of jobs execution. This is because jobs with a stricter deadline must be executed sooner, but due to the resource limitation constraint the scheduler cannot wait for the optimal region to be available. As such, jobs with a shorter deadline tend to be scheduled across multiple regions. In contrast, longer deadline margin allow the scheduler to wait for the optimal region to become available, thus increasing the amount of jobs allocated to optimal locations, even with strict limits on the number of allowed simultaneous workloads.

Our results further show that most jobs are scheduled towards the end of their deadline margin, as Figure \ref{fig:delay-distribution-deadline} clearly shows. This behavior suggests that the scheduler actively leverages the full deadline window to identify time slots with lower carbon intensity, thus maximizing the benefits of time-shifting.

However, the dominating aspect in achieving minimal carbon emissions scheduling ultimately remains the difference in carbon intensity at geographically distributed regions.  While longer deadlines offer more opportunities for optimization, their applicability is workload-dependent. In scenarios where job execution cannot be delayed by one or two days, the gains from time-shifting are necessarily limited.






\section{Conclusion}
\label{conclusion}

In this paper, we propose a Kubernetes-based architecture for the carbon-aware management of cloud workloads to decrease the carbon footprint of cloud computing by applying a time- and space-shifting scheduling approach.

More specifically, we formalize an optimization problem for the placement of Virtual Machines that determines the region and period with minimal carbon intensity. In doing so, we adopt a user-centric approach that does not require information related to the inner workings of data centers -- that is abstracted away by definition in the cloud computing paradigm -- but rather relies on the APIs accessible to cloud computing consumers together with our forecaster component.

The open-source implementation of our architecture is based on Kubernetes as it is the most widely adopted framework for the management of cloud-native resources. In particular, we extend it using Custom Resources Definitions and a Mutating Webhook Configuration for the implementation of custom scheduling logic. Moreover, we focused on supporting user extensibility by adopting the "Policy as Code" paradigm leveraging Open Policy Agent to allow users to express custom constraints on the workload scheduling.

Furthermore, we implemented a forecaster component to predict carbon intensity measurements in a future window at geographically distributed regions. Our proposed model extends TTM, a model for forecasting time series data, and is able to accurately forecast the next 96 hours of carbon intensity data. This enables our system to determine the best time and region with minimal carbon intensity without relying on external third-party services.

We evaluated our proposed architecture using real workload allocation requests under different policies and parameter configurations. The results of our simulations show that our system provides significant reductions in the carbon footprint of cloud computations when compared to a baseline, carbon-agnostic scheduler. Notably, even in the case of a strict scenario with heavy limitations on the eligible schedule regions and resources availability, our system provides up to 13\% less carbon emissions when compared to the baseline -- using internally produced carbon intensity forecasts.

As previously mentioned, one of the key factors in enabling effective and realistic schedules to minimize carbon emissions is the accurate estimation of the limit on the maximum amount of allowed simultaneous jobs at each region. This constraint controls the resource consumption at data centers. The estimation of this parameter is left as future work as it involves researching into the impact of over-scheduling to a single data center, which is beyond the scope of this study.

Another area of refinement that is left as a next step is the implementation of a forecaster for the power consumption profiles of cloud workloads. Indeed, using this information could enable the direct minimization of workload's carbon footprint, instead of searching for the region and window of minimal carbon intensity which is the approach adopted by this study. Moreover, extending the model to consider the dynamic variations during the execution of workloads could enable the implementation of resource scaling techniques such as VMs resizing to further reduce emissions.

Finally, a feature that we are planning to introduce within our system is a reporting system to inform the final user, in a verifiable and trustworthy way, on the carbon emissions savings achieved by adopting our carbon-aware scheduling.

\bibliographystyle{elsarticle-num}
\bibliography{refs}








\end{document}